\documentclass[12pt]{article}
\pdfoutput=1
\usepackage{epsf,amsfonts,amssymb,epsfig,amsmath,graphics,cite}
\usepackage[dvipsnames]{xcolor}
\usepackage{graphicx,subfig}
\usepackage{booktabs}
\usepackage{cancel}
\usepackage{float}
\usepackage{verbatim}
\usepackage{mathrsfs}
\usepackage{empheq}
\usepackage{enumerate}
\usepackage{bbm}
\usepackage{soul}
\usepackage{slashed}
\usepackage[colorlinks=true
,urlcolor=blue
,anchorcolor=blue
,citecolor=blue
,filecolor=blue
,linkcolor=blue
,menucolor=blue
,linktocpage=true
]{hyperref}

\usepackage{tikz}
\usetikzlibrary{cd}
 
\makeatletter

\newcommand{\Rmnum}[1]{\expandafter\@slowromancap\romannumeral #1@}
\makeatother

\newcommand{\nn}{\notag }
\def\be{\begin{equation}}
\def\ee{\end{equation}}

\newcommand{\ii}{\mathrm{i}}

\newcommand{\R}{\mathbb{R}}
\newcommand{\C}{\mathbb{C}}
\newcommand{\Z}{\mathbb{Z}}
\newcommand{\vol}{\mathrm{vol}}

\newcommand{\identity}{\mathbbm{1}}

\newcommand{\mf}[1]{\mathfrak{#1}}

\newcommand{\cA}{\mathcal{A}}

\newcommand{\cC}{\mathcal{C}}

\newcommand{\cF}{\mathcal{F}}

\newcommand{\cH}{\mathcal{H}}
\newcommand{\cI}{\mathcal{I}}

\newcommand{\cK}{\mathcal{K}}
\newcommand{\cL}{\mathcal{L}}

\newcommand{\cN}{\mathcal{N}}
\newcommand{\cO}{\mathcal{O}}

\newcommand{\cQ}{\mathcal Q}

\newcommand{\cY}{\mathcal{Y}}

\renewcommand{\tilde}{\widetilde}

\newcommand{\Itot}{I^{\mathrm{Total}}}

\newcommand{\hook}{\mathbin{\rule[.2ex]{.4em}{.03em}\rule[.2ex]{.03em}{.9ex}}}

\newcommand{\rd}{{\rm d}}

\newcommand{\KKForm}{\alpha}

\newcommand{\FourdSUSYForm}{\eta}
\newcommand{\e}{{\rm e}}

\newcommand{\hooklongrightarrow}{\lhook\joinrel\longrightarrow}

\newcommand{\tp}{\texttt{p}}
\newcommand{\tq}{\texttt{q}}

\numberwithin{equation}{section}       % equation numbers in each section

\begin{document}

\begin{titlepage}

\vfill

\begin{flushright}
Imperial/TP/25/JP/02
\end{flushright}

\vskip 1cm

\begin{center}

{\Large \bf Localizing AlAdS$_5$ black holes \\ 
\vskip 0.2cm
and the SUSY index on $S^1 \times M_3$}

\vskip 1cm
{Jaeha Park}
\vskip 0.1cm

\vskip 1cm

\textit{Abdus Salam Centre for Theoretical Physics,\\
Imperial College London, London SW7 2AZ, UK\\}

%\textit{Blackett Laboratory, Imperial College, \\
%Prince Consort Rd., London, SW7 2AZ, U.K.\\}

\vskip 0.2 cm

\vskip 1cm

\end{center}

\vskip 2 cm

\begin{abstract}
\noindent
We consider complex, supersymmetric, non-extremal Euclidean black holes that are asymptotically locally AdS$_5$, with $S^1 \times M_3$ conformal boundary.
We study field theory backgrounds consisting of various $M_3$, and explicitly construct Killing spinors that are anti-periodic around the Euclidean time circle.
Focussing on elliptically/biaxially squashed three-spheres and Lens spaces, we compute the supersymmetric index of the $\cN=4$ SYM in a Cardy-like limit.
While such black holes have not been constructed for general $M_3$, we demonstrate
that the supersymmetric indices can be recovered from gravity computations using equivariant localization,
by extending the boundary Killing spinors to the bulk.
We show that this involves gluing the black hole geometry with a supersymmetric, horizonless AlAdS$_5$ geometry, chosen such that the Casimir energy is removed from the supersymmetric partition function.

\end{abstract}

\end{titlepage}

\tableofcontents

\newpage
\section{Introduction}\label{app:WCP}

Supersymmetric solutions of supergravity theories with AdS asymptotics have played a key role in holography.
In this case the geometry of the conformal boundary is conformally flat.
Many asymptotically \emph{locally} AdS solutions have also been explicitly constructed \cite{Martelli:2011fu,Martelli:2011fw,Martelli:2013aqa,Cassani:2014zwa,Alday:2014rxa,Alday:2014bta}, as ``gravity duals" of SCFTs on non-conformally flat backgrounds.
In this paper, we consider gravity duals with black hole topology arising as smooth fillings of non-conformally flat boundaries in AdS$_5$/CFT$_4$ holography.

Starting from \cite{BenettiGenolini:2023kxp}, it has recently been shown that various BPS physical observables can be evaluated without solving the supergravity equations, just assuming the solutions exist.
A canonical Killing vector can be constructed as a Killing spinor bilinear, and using the Berline--Vergne--Atiyah--Bott (BVAB) fixed point formula \cite{BV:1982,Atiyah:1984px}, various observables can be computed simply by inputting some topological data. This newly developed formalism, known as equivariant localization in supergravity \cite{BenettiGenolini:2023kxp}, has now been applied to various settings.\footnote{For applications of equivariant localization to internal spaces, see \cite{Martelli:2023oqk,BenettiGenolini:2023yfe,Colombo:2023fhu,BenettiGenolini:2024kyy,Suh:2024asy,Couzens:2024vbn,Hosseini:2025ggq,Hosseini:2025mgf,Couzens:2025nxw}.} For even-dimensional gauged supergravity theories \cite{BenettiGenolini:2024xeo,BenettiGenolini:2024hyd,BenettiGenolini:2024lbj,Couzens:2025ghx}, it was shown that the supergravity on-shell action for all known supergravity solutions can be recovered with ease; moreover, various field theory results, whose gravity duals correspond to \emph{unknown} solutions that are unlikely to ever be found in closed form, were also recovered.

Equivariant localization for $D=5$ gauged supergravity has recently been developed in \cite{BenettiGenolini:2025icr,Colombo:2025ihp}. It was shown that the well-known entropy function of supersymmetric rotating black holes in AdS$_5$ \cite{Hosseini:2017mds,Cabo-Bizet:2018ehj} can be recovered from just topological data, without using the explicit supergravity solutions of \cite{Gutowski:2004ez,Gutowski:2004yv,Chong:2005hr,Kunduri:2006ek}. In this paper, we extend these results to study asymptotically \emph{locally} AdS$_5$ black holes, whose conformal boundary is $S^1 \times M_3$. Known black holes in this class include the numerical solutions of \cite{Cassani:2018mlh,Bombini:2019jhp}, whose conformal boundary $S^1 \times S^3_v$ comprises a ``biaxially squashed" three-sphere, and Lens space black holes with $S^1 \times L(\tp,\tq)$ conformal boundary, that were briefly discussed in \cite{Bobev:2025xan}. We also consider black holes with ``elliptically squashed" conformal boundary $S^1 \times S^3_{\mf{b}_1,\mf{b}_2}$. To the best of our knowledge, supersymmetric black holes filling this conformal boundary have not been previously considered in the literature, and we expect finding such solutions to be very difficult.
Instead of constructing the solutions, our strategy is to explicitly construct only the boundary data, at leading order in the Fefferman-Graham expansion. We demonstrate this is sufficient information to utilise equivariant localization, assuming the boundary Killing vector bilinear $K$, constructed from boundary Killing spinors, extends to the bulk supersymmetric Killing vector, $\cK$.

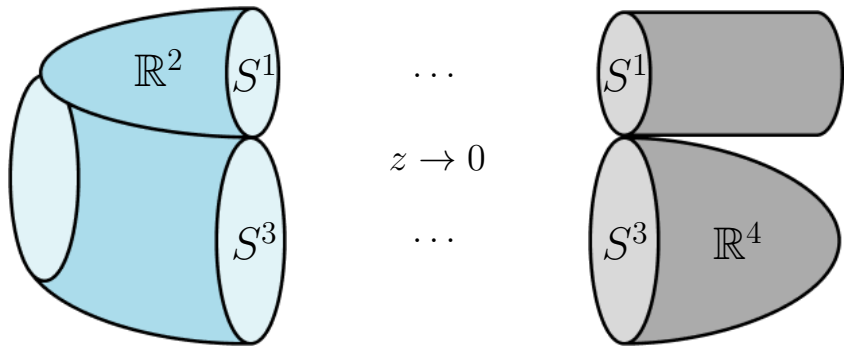
\begin{figure}[H]
        \centering
        \begin{tikzpicture}
            \node[anchor=south east,inner sep=0] at (-2,0)
            {\includegraphics[scale=0.76]{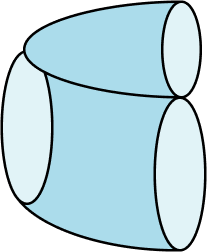}};
            \node[anchor=south west,inner sep=0] at (2,0)
            {\includegraphics[scale=0.76]{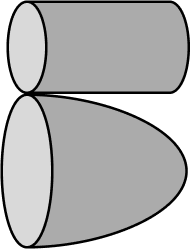}};
            \node[text=black, font=\Large] at (2.5,3.63) {$S^1$};
            \node[text=black, font=\Large] at (2.5,1.4) {$S^3$};
            \node[text=black, font=\Large] at (4,1.4) {$\mathbb{R}^4$};
            \node[text=black, font=\Large] at (-2.43,3.63) {$S^1$};
            \node[text=black, font=\Large] at (-3.7,3.7) {$\R^2$};
            \node[text=black, font=\Large] at (-2.43,1.4) {$S^3$};
            \node[text=black, font=\large] at (0,2.5) {$z \rightarrow 0$};
            \node[text=black, font=\large] at (0,3.6) {$\dots$};
            \node[text=black, font=\large] at (0,1.4) {$\dots$};
        \end{tikzpicture}
  	\centering
	\caption{Schematic description of gluing $M_{(5)} \cong \R^2 \times S^3$ with $N_{(5)} \cong S^1 \times \R^4$, for Euclidean, \emph{squashed}, supersymmetric, non-extremal black holes. The gluing is performed along the common $\partial M_{(5)} = \partial N_{(5)} \cong S^1 \times S^3$ conformal boundary as $z \rightarrow 0$, where $z$ is the Fefferman-Graham coordinate. The $S^3$'s in the figure should be interpreted as squashed three-spheres. Generically, the boundary is \emph{not} conformally flat; both $M_{(5)}$ and $N_{(5)}$ are asymptotically \emph{locally} AdS$_5$. For the Lens space index on $S^1 \times L(\tp, \tq)$, we instead take $M_{(5)} \cong \R^2 \times L(\tp,\tq)$ and $N_{(5)} \cong S^1 \times \R^4/\Z_\tp$.}\label{figure:MglueN}
\end{figure}

For asymptotically AdS black holes, the ``background subtraction" prescription played a key role in the localization computations of \cite{BenettiGenolini:2025icr,Colombo:2025ihp}, where the subtraction geometry was taken to be the AdS vacuum. \emph{A priori}, it is not obvious how this prescription generalises to asymptotically \emph{locally} AdS spacetimes, as they cannot in general be embedded in a suitable ambient spacetime.
Our main result is a holographic prescription that achieves this.
Recall that the supersymmetric index differs from the partition function by $\log Z_{S^1 \times M_3} = - \beta E_{\rm susy} + \log \cI_{S^1 \times M_3}$ \cite{Assel:2014paa,ArabiArdehali:2015iow,ArabiArdehali:2019tdm},
where $E_{\rm susy}$ is the supersymmetric Casimir energy on $S^1 \times M_3$ \cite{Assel:2015nca}.
While $\log Z_{S^1 \times M_3}$ should match a suitably regularised on-shell action of the black hole,
unlike in even-dimensional settings, general supersymmetry preserving counterterms for holographic renormalisation are not known in $D=5$.
By gluing the black hole geometry, $M_{(5)}$, with another gravitational filling of the $S^1 \times M_3$ boundary, $N_{(5)}$, 
it is nevertheless possible to cancel boundary terms that only depend on the intrinsic geometry.
The results of \cite{BenettiGenolini:2016qwm, BenettiGenolini:2016tsn} show that there exist supersymmetric, horizonless AlAdS$_5$ solutions whose on-shell action matches the Casimir energy.
Gluing such solutions with the black hole geometries along their common $S^1 \times M_3$ boundary would result in a closed manifold, as in figure \ref{figure:MglueN}, whose on-shell action matches $\log \cI_{S^1 \times M_3}$.
\\

While the prescription is conceptually straightforward, a necessary requirement is that the (conformal) Killing spinor on $S^1 \times M_3$ extends to both gravitational fillings.
For example, a supersymmetric, horizonless, AlAdS$_5$ geometry with biaxially squashed $S^1 \times S^3_v$ boundary was constructed in \cite{Cassani:2014zwa},
and supersymmetric, extremal black hole solutions with the same conformal boundary has been constructed in \cite{Cassani:2018mlh,Bombini:2019jhp} (see also \cite{Blazquez-Salcedo:2017ghg,Blazquez-Salcedo:2017kig}).
In all of these references, the Killing spinors admitted by the backgrounds are \emph{time-independent}.
To match the supersymmetric index on $S^1 \times S^3_v$, however, one is required to consider a non-extremal deformation of the black hole \cite{Cabo-Bizet:2018ehj}, which has topology $\R^2 \times S^3_v$.
Similar comments apply, \emph{mutatis mutandis}, to other $S^1 \times M_3$ conformal boundaries we consider.
We thus explicitly construct (conformal) Killing spinors on the relevant $S^1_\beta \times M_3$ backgrounds, that are \emph{anti-periodic} around $S^1_\beta = \partial \R^2$, such that they extend to Killing spinors in the bulk.

The asymptotic boundary data we construct is then sufficient to perform the gluing as in figure \ref{figure:MglueN}, and utilise the equivariant localization results of \cite{BenettiGenolini:2025icr}.
For all backgrounds we consider, the gravity computation precisely matches the large $N$ limit of the supersymmetric index on $S^1_\beta \times M_3$.\footnote{Note a computation of the on-shell action of the Lorentzian supersymmetric black holes with $\R \times S^3_v$ boundary was performed in \cite{Cassani:2018mlh,Bombini:2019jhp} using holographic renormalisation in a minimal scheme to cancel divergences. This computation is somewhat formal, as it is done directly using the extremal Lorentzian solution, instead of following the limiting procedure of \cite{Cabo-Bizet:2018ehj}. Accordingly, one should \emph{not} expect the on-shell action of the closed manifold $\R^2 \times S^3_v \, \cup \, (- S^1 \times \R^4_v)$ we find from equivariant localization to agree with the on-shell action computed in \cite{Cassani:2018mlh,Bombini:2019jhp}. An agreement can be made in the microcanonical ensemble instead: the constrained Legendre transform using the on-shell action computed here gives precisely the Bekenstein-Hawking entropy, as shown in (5.18)-(5.19) of \cite{Colombo:2025yqy}.}

The rest of the paper is organised as follows.
In section \ref{fieldtheory}, we explicitly construct supersymmetric backgrounds of the form $S^1_\beta \times M_3$, and compute the supersymmetric index of $SU(N)$ $\cN=4$ SYM on these backgrounds in a Cardy-like limit. 
In section \ref{gravity}, we explain in detail the figure \ref{figure:MglueN} prescription.
Utilising the equivariant localization formalism \cite{BenettiGenolini:2025icr}, an exact match with the field theory computations in section \ref{fieldtheory} is achieved.
We conclude with some discussion in section \ref{discussion}.
Appendix \ref{app:secondsheet} includes definitions of the index using the supercharges admitted by our backgrounds.
Appendix \ref{app:KS} collects useful facts about Killing spinors on squashed three-spheres.

{\bf Note added.} After this work was completed, \cite{Colombo:2025yqy} appeared on the arXiv which has some overlap with our section \ref{gravity}.
Amongst other interesting aspects of their work, they study AlAdS$_5$ solutions with biaxially squashed conformal boundary in their boundary analysis.
From our construction, we are able to explicitly specify asymptotic boundary data for supersymmetric AlAdS$_5$ black holes.

\section{Field theory}\label{fieldtheory}

In this section, we construct rigid supersymmetric backgrounds that arise as the conformal boundary of supersymmetric, non-extremal, asymptocially locally AdS$_5$ black holes.
As shown in \cite{Dumitrescu:2012ha,Klare:2012gn,Closset:2012ru}, $S^1 \times M_3$ is necessarily complex to preserve supersymmetry, and further, when $M_3$ is a Seifert manifold, the background admits two supercharges of opposite R-charge.
In the special case where $M_3$ is a round $S^3$,
the large $N$ limit of the superconformal index \cite{Romelsberger:2005eg,Kinney:2005ej} of $\cN=4$ super-Yang-Mills (SYM) theory has been evaluated using various approaches starting with \cite{Cabo-Bizet:2018ehj,Benini:2018ywd,Choi:2018hmj}, providing a state counting description for the supersymmetric AdS$_5$ black holes of \cite{Gutowski:2004ez,Chong:2005hr} (for a recent review, see \cite{Cassani:2025sim}), matching
the entropy function proposed in \cite{Hosseini:2017mds}
\begin{equation}\label{SCI}
	\log \cI_{S^1_\beta \times S^3} \underset{|\sigma|, |\tau| \rightarrow0}{\simeq} - \ii \pi N^2 \frac{\Delta^1 \Delta^2 \Delta^3}{\sigma \tau} \,, \qquad \Delta^1 + \Delta^2 + \Delta^3 - \sigma - \tau = 1 \,.
\end{equation}
Here, $\Delta_I$ are chemical potentials associated with the holonomies of background gauge fields, and $\sigma$, $\tau$ are complex rotation parameters
that arise as off-diagonal terms in the background metric, often referred to as ``twistings".

A generalisation to the supersymmetric index of the $SU(N)$ $\cN = 4$ SYM on $S^1_\beta \times M_3$ was performed by Honda in \cite{Honda:2019cio}, for general $M_3$, in the Cardy-like limit with the following result:\footnote{Note at this point a saddle has been singled out, namely the $(m,n) = (1,0)$ saddle in the language of \cite{ArabiArdehali:2021nsx}. On the gravity side, when $M_3$ is the round $S^3$, this corresponds to the canonical AdS$_5$ black hole saddle \cite{Cabo-Bizet:2018ehj}, while other $(m,n)$ saddles correspond to orbifolds thereof \cite{Aharony:2021zkr}.}
\begin{align}\label{HondaIndex}
	\log \cI_{S^1_\beta \times M_3} & \underset{\beta \rightarrow 0}{\simeq} -\frac{\pi^3 \ii A_{M_3} (N^2-1)}{\beta^2} \{ m_1 \} \{ m_2 \} (1 - \{ m_1 \} - \{ m_2 \}) \nn \\
	& + \frac{\pi^2 L_{M_3} (N^2-1)}{3\beta} ( \{ m_1 \}^2 + \{ m_2 \}^2 + \{ m_1 \} \{m_2 \} - \{ m_1 \} - \{ m_2 \}  )\,.
\end{align}
Here, $\{ m_{1,2} \}$ are fractional parts of chemical potentials defined as a combination of $\sigma, \tau$ and $\Delta_I$ (see \eqref{defn_m12}).
$A_{M_3}, L_{M_3}$ are integrals of local functionals on $M_3$ given by bosonic fields in the 3d new minimal supergravity multiplet $(h_{\mu\nu}, A_\mu^{(R)}, H, c_\mu)$:
\begin{align}\label{AM3}
	& A_{M_3} = A_{M_3}^{\rm CS} + A_{M_3}^{H} \,, \nn \\
	& A_{M_3}^{\rm CS} = \frac{1}{\pi^2} \int_{M_3} c \wedge {\rd} c \,, \qquad A_{M_3}^{H} = - \frac{\ii}{\pi^2} \int_{M_3} 2 H \, \vol_3 \,,
\end{align}
and
\begin{align}\label{LM3}
	L_{M_3} = \frac{1}{\pi^2} \int_{M_3} {\rd}^3 x \sqrt{h} \left[ - A_\mu ^{(R)} v^\mu + v_\mu v^\mu - \frac{1}{2} H^2 + \frac{1}{4} R \right] \,,
\end{align}
which should be viewed as induced Chern--Simons terms on $M_3$ \cite{DiPietro:2014bca,DiPietro:2016ond}, fixed for a given supersymmetric background on $S^1_\beta \times M_3$.

Starting from suitable ansätze for complex metrics on $S^1_\beta \times M_3$, with non-trivial ``twistings", we explicitly construct (conformal) Killing spinors and new minimal background fields.
By performing a dimensional reduction to derive $(h_{\mu\nu}, A_\mu^{(R)}, H, c_\mu)$, we then compute the supersymmetric index of $SU(N)$ $\cN=4$ SYM on $S^1_\beta \times M_3$ in the Cardy-like limit using \eqref{HondaIndex}, which can be compared with our gravity computation in the large $N$ limit.
After a quick review of the round $S^3$ case in subsection \ref{subsection:S1S3}, we comment on the Lens space index \cite{Benini:2011nc,Razamat:2013opa,Nishioka:2014zpa} in subsection \ref{subsection:LensIndex}.
We then move on to biaxially/elliptically squashed $S^1 \times S^3_v$ and $S^1 \times S^3_{\mf{b}_1,\mf{b}_2}$ backgrounds, respectively, in subsections \ref{subsection:S1S3v} - \ref{subsection:S1M3}.

It should be noted that the 3d effective field theory (EFT) approach \cite{DiPietro:2014bca,DiPietro:2016ond} for the supersymmetric index has already been utilised in the literature \cite{Cassani:2021fyv,ArabiArdehali:2021nsx} to study supersymmetric AdS$_5$ black holes.
In particular, the authors of \cite{Cassani:2021fyv} utilised a real background comprising an elliptically squashed $S^3_{\mf{b}, 1/\mf{b}}$, admitting Killing spinors that are independent of the Euclidean time coordinate. There, the squashing was introduced to capture the effects of twistings by analytic continuation, using holomorphic properties of the supersymmetric index.

Here, we construct a complex supersymmetric $S^1_\beta \times S^3_{\mf{b}_1,\mf{b}_2}$ background, which admits Killing spinors that are \emph{anti-periodic} around $S^1_\beta$.
While the index takes the same form, as expected from holomorphicity, the latter is a necessary requirement for the spin structure to extend to the bulk,
which in turn is necessary for the validity of the equivariant localization computation in section \ref{gravity}.\footnote{The explicit construction of the supersymmetric background presented here, \emph{en passant}, puts the analysis of \cite{Ohmori:2021dzb} on a more concrete footing. The anomaly polynomial in this reference is defined on $\cY_6 \cong \R^2 \times \R^4_{\mf{b}, 1/\mf{b}}$, where $S^1_\beta$ bounds the $\R^2$ disc factor. While the computations go through due to holomorphic properties of the index, formally it is required that the spin structure extends to $\cY_6$. We thank Luigi Tizzano for discussions about this point.}
While the elliptically squashed index is related to the superconformal index (which counts microstates of the known black holes of \cite{Gutowski:2004ez,Chong:2005hr}) by analytic continuations/re-definitions, we thus provide a new interpretation in terms of the counting of microstates for the (yet to be found) AlAdS$_5$ black holes filling in the $S^1_\beta \times S^3_{\mf{b}_1,\mf{b}_2}$ boundary.

On the other hand, the biaxially squashed conformal boundary $S^1 \times S^3_v$ with twistings has not been considered before, and the supersymmetric background constructed here is new.

\subsection{Warm up: \texorpdfstring{$S^1_\beta \times S^3$}{S1×S3} with twistings \texorpdfstring{$\Omega_1$, $\Omega_2$}{Ω1, Ω2}}\label{subsection:S1S3}
As a warm up, we start by revisiting computations in \cite{Cabo-Bizet:2018ehj}, deriving Killing spinors admitted by the following background:
\begin{align}\label{S1S3_metric}
	{\rd} s^2_4 & = {\rd} t_E^2 + {\rd} \vartheta^2 + \sin^2 \vartheta \left({\rd} \varphi_1 -\ii \Omega_1 {\rd} t_E \right)^2 + \cos^2 \vartheta \left({\rd}\varphi_2 - \ii \Omega_2 {\rd} t_E \right)^2 \,, \nn \\
	\cA^R_{\rm bdy} & = \ii \Psi \, {\rd} t_E \,,
\end{align}
where we define $\Psi \equiv \frac{1}{2} ( \Psi_1 + \Psi_2 + \Psi_3 )$, with complex parameters $\Psi_I$ encoded in the background $U(1)^3 \subset SO(6)$ gauge fields specified by the black hole solution.
The coordinates have independent identifications $t_E \sim t_E+\beta$, $\varphi_1 \sim \varphi_1 + 2\pi$, $\varphi_2 \sim \varphi_2+2\pi$. With $\vartheta \in [0,\pi/2]$, this corresponds to a metric on the Hopf surface $S^1 \times S^3$, with ``twistings" parametrised by parameters $\Omega_1, \Omega_2$, that are in general complex. As shown in \cite{Cabo-Bizet:2018ehj}, the background arises at the conformal boundary of complex Euclidean supersymmetric AdS$_5$ black holes.

\subsubsection{Supersymmetry}
To proceed, we use the following the frame
\begin{align}\label{S1S3frame}
	\e^1 & = {\rd} t_E \,, \nn \\
	\e^2 & = \cos(\varphi_1 + \varphi_2) {\rd} \vartheta - \frac{1}{2} \sin(2\vartheta) \sin(\varphi_1 + \varphi_2) \Big( ( {\rd}\varphi_1 -\ii \Omega_1 {\rd}t_E) - ({\rd} \varphi_2 - \ii \Omega_2 {\rd} t_E) \Big) \,, \nn \\
	\e^3 & = \sin(\varphi_1 + \varphi_2) {\rd}\vartheta + \frac{1}{2} \sin(2\vartheta) \cos(\varphi_1 + \varphi_2)  \Big( ( {\rd}\varphi_1 -\ii \Omega_1 {\rd}t_E) - ({\rd} \varphi_2 - \ii \Omega_2 {\rd} t_E) \Big) \,, \nn \\
	\e^4 & = \sin^2 \vartheta \left({\rd} \varphi_1 - \ii \Omega_1 {\rd} t_E \right) + \cos^2\vartheta \left( {\rd}\varphi_2 -\ii \Omega_2 {\rd} t_E \right) \,.
\end{align}
The Killing spinor equation of new minimal supergravity in four-component language is \cite{Sohnius:1981tp,Sohnius:1982fw}:
\begin{equation}\label{KSE_4d_Dirac}
	\left[ \nabla_M - \ii A_M^{\rm nm} \gamma_5 + \ii V_M^{\rm nm} \gamma_5 - \frac{\ii}{2} (V^{\rm nm})^N \gamma_{MN} \gamma_5 \right] \epsilon = 0 \,,
\end{equation}
where the new minimal supergravity gauge field is related to the boundary value of the $D=5$ bulk gauge field by
\begin{equation}
	\cA^R_{\rm bdy} = A^{\rm nm} - \frac{3}{2} V^{\rm nm} = \ii \Psi \, {\rd} t_E \,.
\end{equation}
Following \cite{Cabo-Bizet:2018ehj}, we fix the ambiguity in $A^{\rm nm}, V^{\rm nm}$ by demanding that they only have components along ${\rd} t_E$:
\begin{equation}\label{S1S3_AnmVnm}
	A^{\rm nm} = \ii \left( \Psi - \frac{3}{2} \right) {\rd} t_E \,,\qquad V^{\rm nm} = - \ii \, {\rd} t_E \,.
\end{equation}
The background admits the following Killing spinor\footnote{We use the following representation of gamma matrices:
\begin{align}
	\gamma_1 = \sigma_2 \otimes \mathrm{1}_{2\times2} \,,\qquad \gamma_2 = \sigma_1 \otimes \sigma_1 \,,\qquad \gamma_3 = \sigma_1 \otimes \sigma_2 \,,\qquad \gamma_4 = \sigma_1 \otimes \sigma_3 \,,
\end{align}
with $\gamma_5 = \mathrm{diag}(1,1,-1,-1)$ such that $\gamma_{12345} = \identity$. We take the charge conjugation matrix $\cC = \gamma_2 \gamma_4$.}
\begin{equation}\label{S1S3KSchoice}
	\epsilon = \frac{1}{\sqrt{2}} \begin{pmatrix}
		e^{\frac{t_E}{2} (1-2\Psi + \Omega_1 + \Omega_2)} \\
		0 \\
		0 \\
		e^{-\frac{t_E}{2} (1-2\Psi + \Omega_1 + \Omega_2)} \\
	\end{pmatrix} \,,
\end{equation}
Defining the conjugate spinor
\begin{equation}
	\epsilon^* = \ii \, \cC^{-1} \gamma^1 \epsilon\,,
\end{equation}
the boundary supersymmetric Killing vector is constructed in the following way as a bilinear in the Killing spinors:
\begin{align}\label{S1S3susyKV}
	K & = \epsilon^* \gamma_1 \gamma^M \epsilon \, \partial_M \nn \\
	& = \partial_{t_E} + \ii (\Omega_1 - 1)\partial_{\varphi_1} + \ii (\Omega_2 - 1) \partial_{\varphi_2} \,.
\end{align}
Noting that we are in a frame where $\cL_{\partial_{t_E}} = \partial_{t_E}$ when acting on spinors, we may impose anti-periodic boundary condition $\epsilon(t_E+ \beta) = - \epsilon(t_E)$ by demanding that
\begin{equation}\label{KSE_4d_constraint_S3}
	\beta ( 2\Psi - 1 - \Omega_1 - \Omega_2 ) = 2\pi \ii n \,,\qquad n \ \textrm{odd} \,.
\end{equation}
Defining supersymmetric chemical potentials as
\begin{equation}\label{susychempot_S3}
	\sigma = \frac{\beta(\Omega_1 - 1)}{2\pi \ii} \,,\qquad \tau = \frac{\beta(\Omega_2 - 1)}{2\pi\ii} \,,\qquad \Delta^I = \frac{\beta (\Psi_I - 1)}{2\pi \ii} \,,
\end{equation}
the constraint \eqref{KSE_4d_constraint_S3} can be written as
\begin{equation}\label{SYMconstraint_S3}
	\Delta^1 + \Delta^2 + \Delta^3 - \sigma - \tau = n \,.
\end{equation}

\subsubsection{Superconformal index in the Cardy-like limit}
We now recover the known superconformal index of $\cN=4$ SYM from Honda's formula \eqref{HondaIndex}, setting $M_3$ to be the round $S^3$. The computation extends that of \cite{ArabiArdehali:2021nsx}, which mainly considered the special case $\Omega_1 = \Omega_2$.\footnote{Part of the computation here has been carried out in Appendix D there.}

Dimensional reduction along $t_E$ for 4d multiplets on $S^1_\beta \times M_3$ results in 3d multiplets, as shown in \cite{Closset:2012ru}. The reduction of supersymmetry variations in \cite{Closset:2012ru} assumed that $A^{\rm nm}_{t_E} = V^{\rm nm}_{t_E}$, however this is not satisfied for our background of interest \eqref{S1S3_AnmVnm}.\footnote{Note this is also the assumption in \cite{Cassani:2021fyv} which studies elliptically squashed $S^1 \times S^3_{\mf{b}}$ backgrounds with twistings, see (6.20)-(6.22) there. We lift this assumption in section \ref{subsection:S1M3} (cf. \eqref{AnmVnm_S1S3b}).}
The assumption is lifted in \cite{Assel:2014paa}, and the result is as follows.
Let us write the dimensional reduction ansatz as
\begin{equation}\label{dimredwarp}
	{\rd}s_4^2 = e^{2w} ( {\rd} t_E + c_\mu (x) {\rd}x^\mu)^2 + h_{\mu\nu}(x){\rd}x^\mu {\rd}x^\nu \,.
\end{equation}
Comparing with the background metric \eqref{S1S3_metric}, we find
\begin{align}
	e^{2w} & = 1 - \Omega_1^2 \sin^2 \vartheta - \Omega_2^2 \cos^2 \vartheta \,, \nn \\
	c_\mu {\rd} x^\mu & = - \ii e^{-2w} \left( \Omega_1 \sin^2\vartheta {\rd} \varphi_1 + \Omega_2 \cos^2 \vartheta {\rd} \varphi_2 \right) \,,
\end{align}
as well as
\begin{equation}
	h_{\mu\nu} {\rd}x^\mu {\rd}x^\nu = {\rd}\vartheta^2 + \sin^2 \vartheta {\rd}\varphi_1^2 + \cos^2 \vartheta {\rd}\varphi_2^2 - e^{2w} c^2 \,.
\end{equation}
The 3d new minimal supergravity multiplet is arranged as $(h_{\mu\nu}, A_\mu^{(R)}, H, c_\mu)$, where $h_{\mu\nu}, c_\mu$ follows from the KK ansatz \eqref{dimredwarp}, and $H, A_\mu^{(R)}$ are given by
\begin{equation}\label{H_AR_id}
	H = e^{-w} V^{\rm nm}_{t_E} \,,\qquad A_\mu^{(R)} = A^{\rm nm}_\mu - A^{\rm nm}_{t_E} c_\mu + \frac{1}{2} e^w v_\mu \,,
\end{equation}
where $v$ is the Hodge dual of the graviphoton:
\begin{equation}\label{vHodge}
	v = - \ii \ast_3 {\rd} c \,.
\end{equation}
Given the 4d new minimal supergravity fields \eqref{S1S3_AnmVnm}, it follows that
\begin{equation}\label{H_AR_S1S3}
	H = - \ii e^{-w} \,,\qquad A_\mu^{(R)} = - \ii \left( \frac{1}{2} (\Omega_1 + \Omega_2) - 1 + \frac{\pi \ii n}{\beta} \right) c_\mu + \frac{1}{2} e^w v_\mu \,,
\end{equation}
where we made use of the constraint \eqref{KSE_4d_constraint_S3}.

Collecting the above ingredients, we may now evaluate the local functionals $A_{M_3}, L_{M_3}$ \eqref{AM3}, \eqref{LM3} for the $S^1_\beta \times S^3$ background \eqref{S1S3_metric}, \eqref{S1S3_AnmVnm}. For instance, we have
\begin{equation}
	A_{S^3}^{\rm CS} = \frac{1}{\pi^2} \int_{S^3} c \wedge {\rd} c = - \frac{4\Omega_1 \Omega_2}{(1-\Omega_1^2)(1-\Omega_2^2)} \,,
\end{equation}
as well as
\begin{align}\label{ARvmu}
	\frac{1}{\pi^2} & \int_{S^3} {\rd}^3 x \sqrt{h} \left[ A_\mu^{(R)} v^\mu \right] \nn \\
	& = \frac{2(\Omega_1^2 (1+\Omega_2) + \Omega_2^2 (1+\Omega_1) - 2 \Omega_1 \Omega_2)}{(1-\Omega_1^2)(1-\Omega_2^2)} + \frac{4 \pi \ii n}{\beta} \frac{\Omega_1 \Omega_2}{(1-\Omega_1^2)(1-\Omega_2^2)} \,.
\end{align}
The integrals are rather unwieldy, but they can be evaluated with computer mathematics software. Rather than listing them, we highlight that we are interested only in the asymptotic behaviour in the ``Cardy-like limit" \cite{Choi:2018hmj}, which, in terms of the chemical potentials defined in \eqref{susychempot_S3}, corresponds to the limit $|\sigma|, |\tau| \rightarrow 0$ with fixed $\frac{\sigma}{\tau}$, taken in a way that $\frac{\beta}{\sigma} \ll 1$ and $\frac{\beta}{\tau} \ll 1$. This is a generalisation of the high-temperature limit $\beta \rightarrow 0$ considered in \cite{DiPietro:2014bca,DiPietro:2016ond}.
As explained in \cite{ArabiArdehali:2021nsx}, it is important that the three-dimensional calculations are done on smooth backgrounds, i.e. with finite $\Omega_1, \Omega_2$. We implement this by introducing small parameters $\delta_\sigma = \frac{\beta}{\sigma}$ and $\delta_\tau = \frac{\beta}{\tau}$, such that
\begin{equation}
	\Omega_1 \simeq \frac{2\pi \ii}{\delta_\sigma} \,,\qquad \Omega_2 \simeq \frac{2\pi \ii}{\delta_\tau} \,,
\end{equation}
and taking $\delta_\sigma, \delta_\tau \rightarrow 0$ only at the end of all computations.

It turns out $A_{S^3}^H$ is sub-leading to $A_{S^3}^{\rm CS}$ in the Cardy-like limit, and we have the following asymptotic behaviour of $A_{S^3}$:\footnote{To compare with \cite{Honda:2019cio} note that $A_{M_3} = - A_{M_3}^{\rm there}$.}
\begin{equation}\label{AS3}
	A_{S^3} \underset{\beta \rightarrow 0}{\simeq} \frac{\beta^2}{\pi^2} \frac{1}{\sigma \tau} \,.
\end{equation}
Similarly, contributions from the latter three terms in \eqref{LM3} turn out to be sub-leading to that of \eqref{ARvmu}, and we have the following asymptotic behaviour of $L_{S^3}$:
\begin{equation}\label{LS3}
	L_{S^3} \underset{\beta \rightarrow 0}{\simeq} \frac{\ii \beta}{\pi} \frac{\sigma + \tau}{ \sigma \tau} \,.
\end{equation}
It is worth emphasising that $A_{S^3}, L_{S^3}$ are independent of $\beta$, recall \eqref{susychempot_S3}.
We now insert these results into Honda's formula for the supersymmetric index \eqref{HondaIndex}, noting that $m_{1,2}$ are related to \eqref{susychempot_S3} by \cite{Honda:2019cio}
\begin{equation}\label{defn_m12}
	m_{1,2} = \Delta^{1,2} - \frac{\tau+\sigma}{3} \,,
\end{equation}
satisfying the constraint (setting $n=1$ in \eqref{SYMconstraint_S3})
\begin{equation}\label{SYMconstraint_S3_1}
	\Delta^1 + \Delta^2 + \Delta^3 - \sigma - \tau = 1 \,.
\end{equation}
Honda's formula \eqref{HondaIndex} then implies that the asymptotic behaviour of the superconformal index is given by
\begin{equation}\label{roundSCI}
	\log \cI_{S^1_\beta \times S^3} \underset{|\sigma|, |\tau| \rightarrow 0}{\simeq} - \ii \pi (N^2-1) \frac{\Delta^1 \Delta^2 \Delta^3}{\sigma \tau} \,,
\end{equation}
which indeed matches \eqref{SCI} in the large $N$ limit.

\subsection{Lens space index}\label{subsection:LensIndex}
Given the results in section \ref{subsection:S1S3}, we may draw some immediate conclusions for the Lens space index \cite{Benini:2011nc,Nishioka:2014zpa}.
Recall that a Lens space $L(\tp,\tq)$ is a quotient of $S^3 \subset \C^2$ with complex coordinates $(z_1,z_2)$, where the embedding is $|z_1|^2 + |z_2|^2 = 1$. Given coprime integers $(\tp,\tq)$, the $\Z_{\tp}$ action generated by
\begin{equation}\label{Zpaction}
	(z_1,z_2) \mapsto \left( e^{\frac{2\pi \ii}{\tp}} z_1 , e^{\frac{2\pi \ii \tq}{\tp}} z_2 \right) \,,
\end{equation}
is free, and the resulting quotient space is guaranteed to be a smooth manifold, defined as the Lens space $L(\tp,\tq)$.
Taking
\begin{equation}
	(z_1, z_2) = ( \sin \vartheta e^{\ii \varphi_1}, \cos \vartheta e^{\ii \varphi_2}) \,,
\end{equation}
we may write the same local metric as \eqref{S1S3_metric} on $S^1_\beta \times L(\tp,\tq)$, with $\varphi_1,\varphi_2$ now satisfying orbifold identifications
\begin{equation}\label{Lpqorbiid}
	(\varphi_1,\varphi_2) \sim \left( \varphi_1 + \frac{2\pi}{\tp}, \varphi_2 +\frac{2\pi \tq}{\tp} \right) \,.
\end{equation}
The Killing spinor equations on $S^1_\beta \times L(\tp,\tq)$ are locally the same as those on $S^1_\beta \times S^3$, with the same background fields as in section \ref{subsection:S1S3}. It follows that the functional densities we need to compute $A_{L(\tp,\tq)}, L_{L(\tp,\tq)}$ are the same as the ones we evaluated to obtain \eqref{AS3}, \eqref{LS3}. The only difference comes from the integration measure, which produces an overall factor of $1/\tp$. This observation was made in \cite{DiPietro:2016ond} for $L(\tp,1)$ (see around (2.29) there), but it holds for general $L(\tp,\tq)$.
Consider $S^3$ as a covering space of $L(\tp,\tq) = S^3/\Z_\tp$, with independent coordinate identifications $\varphi_1 \sim \varphi_1 + 2\pi$, $\varphi_2 \sim \varphi_2 + 2\pi$. We can perform the following $SL(2,\Z)$ transformation
\begin{equation}
	\begin{pmatrix}
		\nu \\
		\mu
	\end{pmatrix}
	= \begin{pmatrix}
		1 & 0 \\
		-\tq & 1
	\end{pmatrix}
	\begin{pmatrix}
		\varphi_1 \\
		\varphi_2
	\end{pmatrix} \,,
\end{equation}
with $\Delta \nu = \Delta \mu = 2\pi$ on the covering space. On the quotient space $L(\tp,\tq)$, it follows that \eqref{Lpqorbiid} acts on the $(\mu, \nu)$ coordinates as
\begin{equation}
	(\mu, \nu) \sim \left( \mu , \nu+\frac{2\pi}{\tp}  \right) \,.
\end{equation}
It is then straightforward to see that, for any coprime integers $(\tp,\tq)$, the integration simply produces a factor of $1/\tp$,
\begin{equation}
	A_{L(\tp,\tq)} \underset{\beta \rightarrow 0}{\simeq} \frac{1}{\tp} \frac{\beta^2}{\pi^2} \frac{1}{\sigma \tau} \,, \qquad L_{S^3} \underset{\beta \rightarrow 0}{\simeq} \frac{1}{\tp} \frac{\ii \beta}{\pi} \frac{\sigma + \tau}{ \sigma \tau} \,,
\end{equation}
and the Lens space index in the Cardy-like limit is
\begin{equation}\label{HondaLensIndex}
	\log \cI_{S^1_\beta \times L(\tp,\tq)} \underset{|\sigma|, |\tau| \rightarrow 0}{\simeq} - \frac{\ii \pi (N^2-1)}{\tp} \frac{\Delta^1 \Delta^2 \Delta^3}{\sigma \tau} \,.
\end{equation}

In the large $N$ limit, we expect \eqref{HondaLensIndex} to precisely agree with the on-shell action of AlAdS$_5$ black holes with $L(\tp,\tq)$ horizon topologies, and $S^1 \times L(\tp,\tq)$ conformal boundary.
We note that Lens space black hole solutions have recently been discussed in \cite{Bobev:2025xan}, though the on-shell action was not computed.
In section \ref{gravity}, we demonstrate that \eqref{HondaLensIndex} matches the on-shell action of a closed manifold $\R^2 \times L(\tp,\tq) \cup S^1 \times \R^4/\Z_\tp$ -- see figure \ref{figure:MglueN}.

\subsection{Biaxial squashing \texorpdfstring{$S^1_\beta \times S^3_v$}{S1×S3v} with twistings \texorpdfstring{$\Omega_1$, $\Omega_2$}{Ω1, Ω2}}\label{subsection:S1S3v}
We now consider the following background metric:
\begin{align}\label{S1S3vmetric}
	{\rd}s^2_4 = {\rd} t_E^2 & + {\rd}\vartheta^2 + \sin^2\vartheta \left({\rd} \varphi_1 -\ii \Omega_1 {\rd} t_E \right)^2 + \cos^2\vartheta \left({\rd}\varphi_2 - \ii \Omega_2 {\rd} t_E \right)^2 \nn \\
	& + (v^2-1) \left( \sin^2\vartheta ({\rd}\varphi_1 - \ii \Omega_1 {\rd}t_E) + \cos^2\vartheta ({\rd}\varphi_2 - \ii \Omega_2 {\rd}t_E) \right)^2 \,,
\end{align}
with independent identifications $t_E \sim t_E + \beta$, $\varphi_1 \sim \varphi_1 + 2\pi$, $\varphi_2 \sim \varphi_2 + 2\pi$, with $\vartheta \in [0,\pi/2]$. The spatial part of the metric describes a biaxially squashed $S^3_v$, with squashing parameter $v$ (cf. \eqref{biaxsquashedS3v}).
Setting $\Omega_1 = \Omega_2 = 0$, we recover the direct product metric on $S^1_\beta \times S^3_v$ considered in \cite{Cassani:2014zwa}.\footnote{Note we rescaled the coordinate $t_E$.}
We highlight that the ``twistings" arise from considering the \emph{Euclidean} black hole solution, demanding absence of conical singularities at the horizon. Supersymmetric black hole solutions with biaxially squashed conformal boundary have been constructed numerically in \cite{Cassani:2018mlh,Bombini:2019jhp} (see also \cite{Blazquez-Salcedo:2017ghg,Blazquez-Salcedo:2017kig}) in \emph{Lorentzian} signature, and accordingly their conformal boundary is equivalent to \eqref{S1S3vmetric} without the ``twistings". Supersymmetric, Euclidean, non-extremal deformations of these solutions have not been studied previously in the literature.

\subsubsection{Supersymmetry}
We use the following frame for the metric \eqref{S1S3vmetric}:
\begin{align}\label{S1S3vframe}
	\e^1 & = {\rd} t_E \,, \nn \\
	\e^2 & = \cos(\varphi_1 + \varphi_2) {\rd} \vartheta - \frac{1}{2} \sin(2\vartheta) \sin(\varphi_1 + \varphi_2) \Big( ( {\rd}\varphi_1 -\ii \Omega_1 {\rd}t_E) - ({\rd} \varphi_2 - \ii \Omega_2 {\rd} t_E) \Big) \,, \nn \\
	\e^3 & = \sin(\varphi_1 + \varphi_2) {\rd}\vartheta + \frac{1}{2} \sin(2\vartheta) \cos(\varphi_1 + \varphi_2)  \Big( ( {\rd}\varphi_1 -\ii \Omega_1 {\rd}t_E) - ({\rd} \varphi_2 - \ii \Omega_2 {\rd} t_E) \Big) \,, \nn \\
	\e^4 & = v \left[ \sin^2 \vartheta \left({\rd} \varphi_1 - \ii \Omega_1 {\rd} t_E \right) + \cos^2\vartheta \left( {\rd}\varphi_2 -\ii \Omega_2 {\rd} t_E \right) \right] \,.
\end{align}
We would like to find background field configurations (that is, $A^{\rm nm}$ and $V^{\rm nm}$) that allows for solutions to the Killing spinor equation \eqref{KSE_4d_Dirac}, which we rewrite here:
\begin{equation}
	\left[ \nabla_M - \ii A_M^{\rm nm} \gamma_5 + \ii V_M^{\rm nm} \gamma_5 - \frac{\ii}{2} (V^{\rm nm})^N \gamma_{MN} \gamma_5 \right] \epsilon = 0 \,.
\end{equation}
By choosing
\begin{align}\label{AnmVnm_S1S3v}
	A^{\rm nm} & = \frac{\ii}{2} \left[ \Omega_1 + \Omega_2 - 2 v - \left( 3 - \frac{2}{v} - 2\Psi +  \Omega_1 + \Omega_2 \right) \right] {\rd} t_E \nn \\
	& + (v^2 - 1) \Big[ \ii (\Omega_1 \sin^2\vartheta + \Omega_2 \cos^2\vartheta) {\rd} t_E - \sin^2\vartheta {\rd}\varphi_1 - \cos^2\vartheta {\rd}\varphi_2 \Big] \,, \nn \\
	V^{\rm nm} & = - \ii v \, {\rd} t_E \,,
\end{align}
one can verify that the following is a unit-norm Killing spinor:
\begin{equation}\label{S1S3vKS}
	\epsilon = \frac{1}{\sqrt{2}} \begin{pmatrix}
		\exp \left[ \frac{t_E}{2} \left(3 - \frac{2}{v} - 2\Psi + \Omega_1 + \Omega_2 \right) \right] \\
		0 \\
		0 \\
		\exp \left[ - \frac{t_E}{2} \left(3 - \frac{2}{v} - 2\Psi + \Omega_1 + \Omega_2 \right) \right] \\
	\end{pmatrix} \,.
\end{equation}
Some remarks in order:
\begin{itemize}
	\item We fixed the ambiguity in $A^{\rm nm}, V^{\rm nm}$ by demanding that $V^{\rm nm}$ only has components along ${\rd}t_E$. The background fields were found starting from the assumption that $A^{\rm nm}_{\varphi_1,\varphi_2}$ are given by \eqref{app:S3vAfield}. The $\varphi_1$, $\varphi_2$ components of the Killing spinor equation then fixes $V^{\rm nm}_{t_E}$, and the remaining components are solved by choosing $A^{\rm nm}_{t_E}$ as in \eqref{AnmVnm_S1S3v}.
	\item Sanity check: $A^{\rm nm}$ as given in \eqref{AnmVnm_S1S3v} reduces to \eqref{S1S3_AnmVnm} upon taking $v = 1$.
	\item The gauge field $A^{\rm nm}$ is regular at the two poles of $S^3_v$. As $\vartheta \rightarrow 0$, where ${\rd} \varphi_1$ is not well-defined, we have $A^{\rm nm}_{\varphi_1} \rightarrow 0$, and as $\vartheta \rightarrow \frac{\pi}{2}$, where ${\rd} \varphi_2$ is not well-defined, we have $A^{\rm nm}_{\varphi_2} \rightarrow 0$.
	\item The Lie derivatives $\cL_{\varphi_1}$, $\cL_{\varphi_2}$ acting on individual components of $\epsilon$ are unchanged from the round $S^3$ case, with charges $\pm \frac{1}{2}$. The background admits two supercharges of opposite R-charge.
	\item The following is also a solution:
	\begin{equation}
	\epsilon = \frac{1}{\sqrt{2}} \begin{pmatrix}
		0 \\
		\exp \left[ \frac{t_E}{2} \left(3 - \frac{2}{v} - 2\Psi - \Omega_1 - \Omega_2 \right) \right] \\
		\exp \left[ - \frac{t_E}{2} \left(3 - \frac{2}{v} - 2\Psi - \Omega_1 - \Omega_2 \right)  \right] \\
		0 \\
	\end{pmatrix} \,,
	\end{equation}
	provided for $A^{\rm nm}$ we choose instead
	\begin{align}
	A^{\rm nm} & = - \frac{\ii}{2} \left[ \Omega_1 + \Omega_2 + 2 v + \left( 3 - \frac{2}{v} - 2\Psi - \Omega_1 - \Omega_2 \right) \right] {\rd} t_E \nn \\
	& - (v^2 - 1) \Big[ \ii (\Omega_1 \sin^2\vartheta + \Omega_2 \cos^2\vartheta) {\rd} t_E - \sin^2\vartheta {\rd}\varphi_1 - \cos^2\vartheta {\rd}\varphi_2 \Big] \,.
\end{align}
\end{itemize}
The boundary supersymmetric Killing vector is then given by (recall $\epsilon^* = \ii \cC^{-1} \gamma^1 \epsilon$):
\begin{align}\label{S1S3vsusyKV}
	K & = \epsilon^* \gamma_1 \gamma^M \epsilon \, \partial_M \nn \\
	& = \partial_{t_E} + \ii \left( \Omega_1 - \frac{1}{v} \right) \partial_{\varphi_1} + \ii \left( \Omega_2 - \frac{1}{v} \right) \partial_{\varphi_2} \,.
\end{align}
The frame \eqref{S1S3vframe} is such that $\cL_{\partial_{t_E}} \epsilon = \partial_{t_E} \epsilon$. It follows that anti-periodic boundary condition on the Killing spinor \eqref{S1S3vKS}, given by $\epsilon(t_E+\beta) = -\epsilon(t_E)$, requires:
\begin{equation}\label{KSE_4d_constraint_S3v}
	\beta \left( 2\Psi - 3 + \frac{2}{v} - \Omega_1 - \Omega_2 \right)  = 2\pi \ii n \,,\qquad n \ \textrm{odd} \,.
\end{equation}
We now define ``biaxially squashed" supersymmetric chemical potentials as
\begin{equation}\label{susychempot_S3v}
	\sigma_v = \frac{\beta}{2\pi\ii} \left( \Omega_1 - \frac{1}{v} \right) \,,\qquad \tau_v =  \frac{\beta}{2\pi\ii} \left( \Omega_2 - \frac{1}{v} \right) \,,\qquad \Delta^I_v = \frac{\beta ( \Psi_I - 1 )}{2\pi\ii} \,,
\end{equation}
such that the constraint reads
\begin{equation}\label{SYMconstraint_S3v}
	\Delta^1_v + \Delta^2_v + \Delta^3_v - \sigma_v - \tau_v = n \,,\qquad n \ \textrm{odd} \,.
\end{equation}

\subsubsection{Biaxially squashed index in the Cardy-like limit}
We now perform the KK reduction along $t_E$, starting from the metric \eqref{S1S3vmetric}, identify 3d new minimal supergravity fields, and evaluate $A_{M_3}, L_{M_3}$ given by \eqref{AM3}, \eqref{LM3} in the Cardy-like limit.

As before, let us write the dimensional reduction ansatz as
\begin{equation}
	{\rd}s_4^2 = e^{2w} ( {\rd} t_E + c_\mu (x) {\rd}x^\mu)^2 + h_{\mu\nu}(x){\rd}x^\mu {\rd}x^\nu \,.
\end{equation}
Comparing with the background metric \eqref{S1S3vmetric}, we find
\begin{align}
	e^{2w} = 1 - \Omega_1^2 \sin^2 \vartheta- \Omega_2^2 \cos^2 \vartheta - (v^2 - 1) ( \Omega_1 \sin^2\vartheta + \Omega_2 \cos^2 \vartheta)^2 \,, \nn \\
	c_\mu {\rd} x^\mu = - \ii e^{-2w} \Big[ \sin^2\vartheta \left( \Omega_1 + (v^2-1) (\Omega_1 \sin^2\vartheta + \Omega_2 \cos^2\vartheta) \right) {\rd} \varphi_1 \nn \\
	+ \cos^2\vartheta \left( \Omega_2 + (v^2-1) (\Omega_1 \sin^2\vartheta + \Omega_2 \cos^2\vartheta) \right) {\rd} \varphi_2 \Big] \,,
\end{align}
as well as
\begin{align}
	& h_{\mu\nu} {\rd}x^\mu {\rd}x^\nu \nn \\
	& = {\rd}\vartheta^2 + \sin^2\vartheta {\rd}\varphi_1^2 + \cos^2\vartheta {\rd}\varphi_2^2 + (v^2-1) \left( \sin^2\vartheta {\rd}\varphi_1 + \cos^2\vartheta {\rd}\varphi_2 \right)^2 - e^{2w} c^2 \,.
\end{align}
From the dimensional reduction \cite{Assel:2014paa}, 3d new minimal supergravity fields are identified via \eqref{H_AR_id}-\eqref{vHodge}. Given \eqref{AnmVnm_S1S3v}, we have (cf. \eqref{H_AR_S1S3})
\begin{align}\label{H_AR_S1S3v}
	H & = - \ii v e^{-w} \,, \nn \\
	A_\mu^{(R)} & = - (v^2 - 1) \left( \sin^2\vartheta {\rd} \varphi_1 + \cos^2\vartheta {\rd} \varphi_2 \right) + \frac{1}{2} e^w v_\mu \nn \\
	& - \ii \left[ \frac{1}{2} ( \Omega_1+ \Omega_2 ) - v + \frac{\pi\ii n}{\beta} + (v^2 - 1)(\Omega_1 \sin^2\vartheta + \Omega_2 \cos^2 \vartheta) \right] c_\mu 
\end{align}
where we have used the constraint \eqref{KSE_4d_constraint_S3v}.

Collecting the above ingredients, we may now evaluate the local functionals $A_{M_3}, L_{M_3}$ \eqref{AM3}, \eqref{LM3} for the $S^1_\beta \times S^3_v$ background \eqref{S1S3vmetric}, \eqref{AnmVnm_S1S3v}. We note that the integrals were significantly more complicated compared to the round case, and to the elliptically squashed case that we discuss in subsection \ref{subsection:S1M3}. We did not succeed in evaluating them before taking the Cardy-like limit.
Nevertheless, it was possible to first take the Cardy-like limit within the integrand and integrate the series expansion.
As before, with the biaxially squashed chemical potentials \eqref{susychempot_S3v}, the Cardy-like limit is taken by introducing small parameters $\delta_{\sigma_v} = \frac{\beta}{\sigma_v}$ and $\delta_{\tau_v} = \frac{\beta}{\tau_v}$, such that
\begin{equation}
	\Omega_1 \simeq \frac{2\pi \ii}{\delta_{\sigma_v}} \,,\qquad \Omega_2 \simeq \frac{2\pi \ii}{\delta_{\tau_v}} \,,
\end{equation}
and taking $\delta_{\sigma_v}, \delta_{\tau_v} \rightarrow 0$ with fixed $\frac{\sigma_v}{\tau_v}$. We find
\begin{equation}\label{AS3vLS3v}
	A_{S^3_v} \underset{\beta \rightarrow 0}{\simeq} \frac{\beta^2}{\pi^2} \frac{1}{\sigma_v \tau_v} \,, \qquad L_{S^3_v} \underset{\beta \rightarrow 0}{\simeq} \frac{\ii \beta}{\pi} \frac{\sigma_v + \tau_v}{\sigma_v \tau_v} \,.
\end{equation}
Substituting these into Honda's formula \eqref{HondaIndex}, with ``biaxially squashed" chemical potentials satisfying the constraint (cf. \eqref{SYMconstraint_S3v})
\begin{equation}\label{SYMconstraint_S3v_1}
	\Delta_{v}^1 + \Delta_{v}^2 + \Delta_{v}^3 - \sigma_{v} - \tau_{v} = 1 \,,
\end{equation}
we conclude that the biaxially squashed index in the Cardy-like limit is given by
\begin{equation}\label{HondaSquashedIndex}
	\log \cI_{S^1_\beta \times S^3_{v}} \underset{|\sigma_v|, |\tau_v| \rightarrow 0}{\simeq} - \ii \pi (N^2-1) \frac{\Delta_{v}^1\Delta_{v}^2\Delta_{v}^3}{\sigma_v \tau_v} \,.
\end{equation}
While the result takes the same form as the round $S^3$ case, \eqref{roundSCI}, we highlight that the definitions of the chemical potentials are different, depending non-trivially on the squashing parameter \eqref{susychempot_S3v}.

\subsection{Elliptical squashing \texorpdfstring{$S^1_\beta \times S^3_{\mf{b}_1,\mf{b}_2}}{S1×S3b1b2}$ with twistings \texorpdfstring{$\Omega_1$, $\Omega_2$}{Ω1, Ω2}}\label{subsection:S1M3}
We now consider the following background metric:
\begin{align}\label{S1S3bmetric}
	{\rd} s^2_4 & = {\rd} t_E^2 + f(\vartheta)^2 {\rd} \vartheta^2 + \frac{\sin^2 \vartheta}{\mf{b}_1^2} \left({\rd} \varphi_1 -\ii \Omega_1 {\rd} t_E \right)^2 + \frac{\cos^2 \vartheta}{\mf{b}_2^2} \left({\rd}\varphi_2 - \ii \Omega_2 {\rd} t_E \right)^2 \,,
\end{align}
with independent coordinate identifications $t_E \sim t_E + \beta$, $\varphi_1 \sim \varphi_1 + 2\pi$, $\varphi_2 \sim \varphi_2 + 2\pi$, as well as $\vartheta \in [0, \pi/2 ]$. The real parameters $\mf{b}_1, \mf{b}_2$ control the ``squashings".
As in section \ref{subsection:S1S3}, the ``twisting" parameters $\Omega_1, \Omega_2$ are \emph{complex}.
If we set
\begin{equation}\label{f_vartheta}
	f(\vartheta) \equiv \sqrt{\frac{\cos^2\vartheta}{\mf{b}_1^2} + \frac{\sin^2\vartheta}{\mf{b}_2^2}} \,,
\end{equation}
the spatial part of the metric is equivalent to the elliptically squashed three-sphere of \cite{Hama:2011ea}.
We will not specify the function $f(\vartheta)$ and thus allow for arbitrary metrics on $S^3_{\mf{b}_1, \mf{b}_2}$, following observations in e.g. \cite{Martelli:2011fu} that physical observables do not depend on the details of $f(\vartheta)$.

Note a similar, real background was considered in \cite{Cassani:2021fyv}, restricting to pure imaginary $\Omega_1, \Omega_2$ and \eqref{f_vartheta}, that admits Killing spinors that are periodic around the Euclidean time circle. Here, we construct anti-periodic Killing spinors, that naturally extend to Killing spinors in the bulk, for the equivariant localization computation in section \ref{gravity}.
As the 4d supersymmetric partition function is a holomorphic function of the fugacities \cite{Closset:2013vra}, we expect the resulting expression of the index to take the same form. From a 3d EFT point of view, however, it is non-trivial that the Chern--Simons terms evaluated on our choice of background fields yield the same result.

\subsubsection{Supersymmetry}
We use the following frame for the metric \eqref{S1S3bmetric}:
\begin{align}\label{S1S3bframe}
	\e^1 & = {\rd} t_E \,, \nn \\
	\e^2 & = \cos(\varphi_1 + \varphi_2) f(\vartheta) {\rd} \vartheta \nn \\
	& - \frac{1}{2} \sin(2\vartheta) \sin(\varphi_1 + \varphi_2) \left( \frac{1}{\mf{b}_1} ( {\rd}\varphi_1 -\ii \Omega_1 {\rd}t_E) - \frac{1}{\mf{b}_2} ({\rd} \varphi_2 - \ii \Omega_2 {\rd} t_E) \right) \,, \nn \\
	\e^3 & = \sin(\varphi_1 + \varphi_2) f(\vartheta) {\rd}\vartheta \nn \\
	& + \frac{1}{2} \sin(2\vartheta) \cos(\varphi_1 + \varphi_2)  \left( \frac{1}{\mf{b}_1} ( {\rd}\varphi_1 -\ii \Omega_1 {\rd}t_E) - \frac{1}{\mf{b}_2} ({\rd} \varphi_2 - \ii \Omega_2 {\rd} t_E) \right) \,, \nn \\
	\e^4 & = \frac{\sin^2 \vartheta}{\mf{b}_1} \left({\rd} \varphi_1 - \ii \Omega_1 {\rd} t_E \right) + \frac{\cos^2\vartheta}{\mf{b}_2} \left( {\rd}\varphi_2 -\ii \Omega_2 {\rd} t_E \right) \,.
\end{align}
We would like to find background field configurations (that is, $A^{\rm nm}$ and $V^{\rm nm}$) that allows for solutions to the Killing spinor equation \eqref{KSE_4d_Dirac}, which we rewrite here:
\begin{equation}
	\left[ \nabla_M - \ii A_M^{\rm nm} \gamma_5 + \ii V_M^{\rm nm} \gamma_5 - \frac{\ii}{2} (V^{\rm nm})^N \gamma_{MN} \gamma_5 \right] \epsilon = 0 \,.
\end{equation}
By choosing
\begin{align}\label{AnmVnm_S1S3b}
	A^{\rm nm} & = \frac{\ii}{2} \left[ \frac{1}{f(\vartheta)} \left( \frac{\Omega_1}{\mf{b}_1} + \frac{\Omega_2}{\mf{b}_2} - 2  \right) - \left( 3 - \mf{b}_1 - \mf{b}_2 - 2\Psi +  \Omega_1 + \Omega_2 \right) \right] {\rd} t_E  \nn \\
	& + \frac{1}{2} \left( 1- \frac{1}{f(\vartheta) \mf{b}_1} \right) {\rd}\varphi_1 + \frac{1}{2} \left( 1- \frac{1}{f(\vartheta) \mf{b}_2} \right) {\rd}\varphi_2 \,, \nn \\
	V^{\rm nm} & = - \frac{\ii}{f(\vartheta)} {\rd} t_E \,,
\end{align}
one can verify that the following is a unit-norm Killing spinor (cf. \eqref{S1S3KSchoice}):
\begin{equation}\label{S1S3bKS}
	\epsilon = \frac{1}{\sqrt{2}} \begin{pmatrix}
		\exp \left[ \frac{t_E}{2} \left(3 - \mf{b}_1 - \mf{b}_2 - 2\Psi + \Omega_1 + \Omega_2 \right) \right] \\
		0 \\
		0 \\
		\exp \left[ - \frac{t_E}{2} \left(3 - \mf{b}_1 - \mf{b}_2 - 2\Psi + \Omega_1 + \Omega_2 \right) \right] \\
	\end{pmatrix} \,.
\end{equation}
Some remarks in order:
\begin{itemize}
	\item We fixed the ambiguity in $A^{\rm nm}, V^{\rm nm}$ by demanding that $V^{\rm nm}$ only has components along ${\rd} t_E$. The background fields were found starting from the assumption that $A^{\rm nm}_{\varphi_1,\varphi_2}$ are given by \eqref{app:S3bAfield}. The $\varphi_1$, $\varphi_2$ components of the Killing spinor equation then fixes $V^{\rm nm}_{t_E}$, and the remaining components are solved by choosing $A^{\rm nm}_{t_E}$ as in \eqref{AnmVnm_S1S3b}.
	\item Sanity check: $A^{\rm nm}$ as given in \eqref{AnmVnm_S1S3b} reduces to that as given in \eqref{S1S3_AnmVnm} upon taking $\mf{b}_1 = \mf{b}_2 = 1$, given that $f(\vartheta)$ reduces to 1.
	\item The gauge field $A^{\rm nm}$ is regular at the two poles of $S^3_{\mf{b}_1,\mf{b}_2}$, provided $f(\vartheta)$ is given by \eqref{f_vartheta}. As $\vartheta \rightarrow 0$, where ${\rd} \varphi_1$ is not well-defined, we have $A^{\rm nm}_{\varphi_1} \rightarrow 0$, and as $\vartheta \rightarrow \frac{\pi}{2}$, where ${\rd} \varphi_2$ is not well-defined, we have $A^{\rm nm}_{\varphi_2} \rightarrow 0$.
	\item The Lie derivatives $\cL_{\varphi_1}$, $\cL_{\varphi_2}$ acting on individual components of $\epsilon$ are unchanged from the round $S^3$ case, with charges $\pm \frac{1}{2}$. The background admits two supercharges of opposite R-charge.
	\item The following is also a solution:
	\begin{equation}
	\epsilon = \frac{1}{\sqrt{2}} \begin{pmatrix}
		0 \\
		\exp \left[ \frac{t_E}{2} \left(3 - \mf{b}_1 - \mf{b}_2 - 2\Psi - \Omega_1 - \Omega_2 \right) \right] \\
		\exp \left[ - \frac{t_E}{2} \left(3 - \mf{b}_1 - \mf{b}_2 - 2\Psi - \Omega_1 - \Omega_2 \right) \right] \\
		0 \\
	\end{pmatrix} \,,
	\end{equation}
	provided that for $A^{\rm nm}$ we choose instead
	\begin{align}
	A^{\rm nm} & = - \frac{\ii}{2} \left[ \frac{1}{f(\vartheta)} \left( \frac{\Omega_1}{\mf{b}_1} + \frac{\Omega_2}{\mf{b}_2} + 2  \right) + \left( 3 - \mf{b}_1 - \mf{b}_2 - 2\Psi -  \Omega_1 - \Omega_2 \right) \right] {\rd} t_E \nn \\
	& - \frac{1}{2} \left( 1- \frac{1}{f(\vartheta) \mf{b}_1} \right) {\rd}\varphi_1 - \frac{1}{2} \left( 1- \frac{1}{f(\vartheta) \mf{b}_2} \right) {\rd}\varphi_2 \,.
\end{align}
\end{itemize}
The boundary supersymmetric Killing vector is then given by (recall $\epsilon^* = \ii \cC^{-1} \gamma^1 \epsilon$):
\begin{align}\label{S1S3bsusyKV}
	K & = \epsilon^* \gamma_1 \gamma^M \epsilon \, \partial_M \nn \\
	& = \partial_{t_E} + \ii (\Omega_1 - \mf{b}_1)\partial_{\varphi_1} + \ii (\Omega_2 - \mf{b}_2) \partial_{\varphi_2} \,.
\end{align}
The frame \eqref{S1S3bframe} is such that $\cL_{\partial_{t_E}} \epsilon = \partial_{t_E} \epsilon$. It follows that anti-periodic boundary condition on the Killing spinor \eqref{S1S3bKS}, given by $\epsilon(t_E+\beta) = -\epsilon(t_E)$, requires (cf. \eqref{KSE_4d_constraint_S3}):
\begin{equation}\label{KSE_4d_constraint_S3b}
	\beta \left( 2\Psi - 3 + \mf{b}_1 + \mf{b}_2 - \Omega_1 - \Omega_2 \right)  = 2\pi \ii n \,,\qquad n \ \textrm{odd} \,.
\end{equation}
We can now define ``elliptically squashed" supersymmetric chemical potentials as
\begin{equation}\label{susychempot_S3b}
	\sigma_{\mf{b}} = \frac{\beta(\Omega_1 - \mf{b}_1)}{2\pi\ii} \,,\qquad \tau_{\mf{b}} = \frac{\beta(\Omega_2 - \mf{b}_2)}{2\pi\ii} \,,\qquad \Delta^I_{\mf{b}} = \frac{\beta( \Psi_I - 1 )}{2\pi\ii} \,,
\end{equation}
and the constraint \eqref{KSE_4d_constraint_S3b} is given by
\begin{equation}\label{SYMconstraint_S3b}
	\Delta^1_{\mf{b}} + \Delta^2_{\mf{b}} + \Delta^3_{\mf{b}} - \sigma_{\mf{b}}- \tau_{\mf{b}} = n \,,\qquad n \ \textrm{odd} \,.
\end{equation}

\subsubsection{Elliptically squashed index in the Cardy-like limit}

We now perform the KK reduction along $t_E$, starting from the metric \eqref{S1S3bmetric}, identify 3d new minimal supergravity fields, and evaluate $A_{M_3}, L_{M_3}$ given by \eqref{AM3}, \eqref{LM3} in the Cardy-like limit. A similar computation was carried out in section 6 of \cite{Cassani:2021fyv}, though with reality assumptions on the 4d metric, as well as assuming $A^{\rm nm}_{t_E} = V^{\rm nm}_{t_E}$. We are required to lift both assumptions, as $\Omega_1, \Omega_2$ are complex, and the latter assumption does not hold for our background of interest (recall \eqref{AnmVnm_S1S3b}). The result is as follows.

As before, let us write the dimensional reduction ansatz as
\begin{equation}
	{\rd}s_4^2 = e^{2w} ( {\rd} t_E + c_\mu (x) {\rd}x^\mu)^2 + h_{\mu\nu}(x){\rd}x^\mu {\rd}x^\nu \,.
\end{equation}
Comparing with the background metric \eqref{S1S3bmetric}, we find
\begin{align}
	e^{2w} & = 1 - \frac{\Omega_1^2 \sin^2 \vartheta}{\mf{b}_1^2} - \frac{\Omega_2^2 \cos^2 \vartheta}{\mf{b}_2^2} \,, \nn \\
	c_\mu {\rd} x^\mu & = - \ii e^{-2w} \left( \frac{\Omega_1 \sin^2\vartheta}{\mf{b}_1^2} {\rd} \varphi_1 + \frac{\Omega_2 \cos^2 \vartheta}{\mf{b}_2^2} {\rd} \varphi_2 \right) \,,
\end{align}
as well as
\begin{equation}
	h_{\mu\nu} {\rd}x^\mu {\rd}x^\nu = f(\vartheta)^2 {\rd}\vartheta^2 + \frac{\sin^2 \vartheta}{\mf{b}_1^2} {\rd}\varphi_1^2 + \frac{\cos^2 \vartheta}{\mf{b}_2^2} {\rd}\varphi_2^2 - e^{2w} c^2 \,.
\end{equation}
From the dimensional reduction \cite{Assel:2014paa}, 3d new minimal supergravity fields are identified via \eqref{H_AR_id}-\eqref{vHodge}. Given \eqref{AnmVnm_S1S3b}, we have (cf. \eqref{H_AR_S1S3})
\begin{align}\label{H_AR_S1S3b}
	H & = - \frac{\ii}{f(\vartheta)} e^{-w} \,, \nn \\
	A_\mu^{(R)} & = \frac{1}{2} \left( 1- \frac{1}{f(\vartheta) \mf{b}_1} \right) {\rd} \varphi_1 + \frac{1}{2} \left( 1- \frac{1}{f(\vartheta) \mf{b}_2} \right) {\rd} \varphi_2 \nn \\
	& - \ii \left[ \frac{1}{2f(\vartheta)} \left( \frac{\Omega_1}{\mf{b}_1} + \frac{\Omega_2}{\mf{b}_2} - 2  \right) + \frac{\pi \ii n}{\beta} \right] c_\mu + \frac{1}{2} e^w v_\mu 
\end{align}
where we have used the constraint \eqref{KSE_4d_constraint_S3b}.

Collecting the above ingredients, we may now compute $A_{S^3_{\mf{b}_1,\mf{b}_2}}$ and $L_{S^3_{\mf{b}_1,\mf{b}_2}}$ given by \eqref{AM3}, \eqref{LM3}. We note that in evaluating
\begin{equation}
	A_{S^3_{\mf{b}_1,\mf{b}_2}}^{\rm CS} = \frac{1}{\pi^2} \int_{S^3} c \wedge {\rd} c = - \frac{4\Omega_1 \Omega_2}{(\mf{b}_1^2-\Omega_1^2)(\mf{b}_2^2-\Omega_2^2)} \,,
\end{equation}
the dependence on $f(\vartheta)$ explicitly cancels out in the integrand and $A_{S^3_{\mf{b}_1,\mf{b}_2}}^{\rm CS}$ does not depend on the function at all.
With the modified chemical potentials \eqref{susychempot_S3b}, we now take the ``Cardy-like limit" by introducing small parameters $\delta_{\sigma_{\mf{b}}} = \frac{\beta}{\sigma_{\mf{b}}}$ and $\delta_{\tau_{\mf{b}}} = \frac{\beta}{\tau_{\mf{b}}}$, such that
\begin{equation}
	\Omega_1 \simeq \frac{2\pi \ii}{\delta_{\sigma_{\mf{b}}}} \,,\qquad \Omega_2 \simeq \frac{2\pi \ii}{\delta_{\tau_{\mf{b}}}} \,,
\end{equation}
and taking $\delta_{\sigma_{\mf{b}}}, \delta_{\tau_{\mf{b}}} \rightarrow 0$ only at the end of all computations, with fixed $\frac{ \sigma_{\mf{b}} }{ \tau_{\mf{b}} }$.
As observed in the round $S^3$ analysis in section \ref{subsection:S1S3}, $A^H_{S^3_{\mf{b}_1,\mf{b}_2}}$ is sub-leading to $A^{\rm CS}_{S^3_{\mf{b}_1,\mf{b}_2}}$ in the modified Cardy-like limit, and we have
\begin{equation}\label{AS3b}
	A_{S^3_{\mf{b}_1,\mf{b}_2}} \underset{\beta \rightarrow 0}{\simeq} \frac{\beta^2}{\pi^2} \frac{1}{\sigma_{\mf{b}} \tau_{\mf{b}}} \,.
\end{equation}
Similarly, the leading asymptotic behaviour of $L_{S^3_{\mf{b}_1,\mf{b}_2}}$ in \eqref{LM3} comes from the $A_\mu^{(R)} v^\mu$ term, and we have
\begin{equation}\label{LS3b}
	L_{S^3_{\mf{b}_1,\mf{b}_2}} \underset{\beta \rightarrow 0}{\simeq} \frac{\ii \beta}{\pi} \frac{\sigma_{\mf{b}} + \tau_{\mf{b}}}{\sigma_{\mf{b}} \tau_{\mf{b}}} \,.
\end{equation}
Note to evaluate the integrals required to compute $L_{S^3_{\mf{b}_1,\mf{b}_2}}$, we had to specify the function $f(\vartheta)$ as in \eqref{f_vartheta}, but it is likely that the dependence drops out in the Cardy-like limit. Indeed, if one takes the limit first (as opposed to taking the limit after the integration), the terms involving $f(\vartheta)$ drop out, and we recover the same result, that is, \eqref{LS3b}.
Substituting these into Honda's formula \eqref{HondaIndex}, with ``elliptically squashed" chemical potentials satisfying the constraint (setting $n=1$ in \eqref{SYMconstraint_S3b})
\begin{equation}\label{SYMconstraint_S3b_1}
	\Delta_{\mf{b}}^1 + \Delta_{\mf{b}}^2 + \Delta_{\mf{b}}^3 - \sigma_{\mf{b}} - \tau_{\mf{b}} = 1 \,,
\end{equation}
we conclude that the elliptically squashed index in the Cardy-like limit is given by
\begin{equation}\label{HondaSquashedIndexb}
	\log \cI_{S^1_\beta \times S^3_{\mf{b}_1,\mf{b}_2}} \underset{|\sigma_{\mf{b}}|, |\tau_{\mf{b}}| \rightarrow 0}{\simeq} - \ii \pi (N^2-1) \frac{\Delta_{\mf{b}}^1\Delta_{\mf{b}}^2\Delta_{\mf{b}}^3}{\sigma_{\mf{b}} \tau_{\mf{b}}} \,.
\end{equation}

It is interesting to compare this to the ``second sheet" result obtained in \cite{Cassani:2021fyv}, where the Chern--Simons terms were evaluated on a real background.
Recall for this comparison that we computed the index of the $SU(N)$ $\cN=4$ SYM, for which $a = c = \frac{N^2-1}{4}$. Setting $\Delta^1_{\mf{b}} = \Delta^2_{\mf{b}} = \Delta^3_{\mf{b}} \equiv \frac{2}{3} \Delta_{\mf{b}}$, and restricting $\Omega_1, \Omega_2$ to be purely imaginary, we may identify
\begin{equation}\label{CKid}
	\omega_1^{\rm there} \leftrightarrow - 2\pi \ii \sigma_{\mf{b}} \,,\qquad \omega_2^{\rm there} \leftrightarrow - 2\pi \ii \tau_{\mf{b}} \,,
\end{equation}
and one finds that \eqref{HondaSquashedIndexb} matches \cite[(6.36)]{Cassani:2021fyv}, with $n_0^{\rm there} = -1$.
The fact that the two results agree strongly suggests that the two supersymmetric backgrounds, that is, (6.2), (6.6) there and \eqref{S1S3bmetric}, \eqref{AnmVnm_S1S3b} here, are related by a supersymmetry-preserving deformation that leaves the partition function invariant.\footnote{For related discussions, see around (3.13) of \cite{Cabo-Bizet:2021jar}, as well as \cite[sect. 2]{BenettiGenolini:2023rkq}.}
We note that the ``shift" in the chemical potentials considered in \cite{Cassani:2021fyv} to go to the ``second sheet" is precisely such that the constraint \eqref{SYMconstraint_S3b_1} is satisfied. Here, we showed that the constraint can be realised from the periodicity condition \eqref{KSE_4d_constraint_S3b} without having to perform such a ``shift", by directly constructing a complex supersymmetric background; this generalises the Killing spinors in \cite{Cabo-Bizet:2018ehj} to incorporate squashing parameters $\mf{b}_1, \mf{b}_2$. In the next section, we will show that the result \eqref{HondaSquashedIndexb} can be associated with the on-shell action of a non-extremal, supersymmetric, asymptotically locally AdS$_5$ black hole filling in the $S^1_\beta \times S^3_{\mf{b}_1,\mf{b}_2}$ boundary, providing a new interpretation for the results of \cite{Cassani:2021fyv}.

\section{Gravity}\label{gravity}
We now consider complex, non-extremal deformations of supersymmetric extremal black holes with topology $\R^2 \times M_3$. While such solutions are known for $M_3 = S^3$ \cite{Cabo-Bizet:2018ehj}, there are no known non-extremal supersymmetric deformations for the numerical black hole solutions of \cite{Cassani:2018mlh,Bombini:2019jhp}, with biaxially squashed $S^1_\beta \times S^3_v$ boundary. Even less is known for the elliptically squashed $S^1_\beta \times S^3_{\mf{b}_1, \mf{b}_2}$ boundary: to the best of our knowledge, supersymmetric black holes filling this boundary have never been constructed.
Remarkably, we find that supersymmetric indices on $S^1_\beta \times M_3$ can be recovered using the equivariant localization formalism developed in \cite{BenettiGenolini:2025icr}, without explicit knowledge of the solutions.
The only ingredients are topological data and the R-symmetry Killing vector, $\cK$, where the latter is obtained as an extension of the (conformal) Killing vector $K$ formed as a bilinear in the boundary Killing spinors (similar to \cite{Anderson:2007jpe}).

In the formalism of \cite{BenettiGenolini:2025icr}, in addition to $\cK$, it is assumed that the Euclidean supersymmetric solution on $M_{(5)}$ admits an additional Killing vector $\ell$.
Performing a dimensional reduction along $\ell$, the on-shell action of the resulting $D=4$, $\cN=2$ supergravity localises into a sum over fixed points of $\xi = \pi_*(\cK)$  \cite{BenettiGenolini:2024lbj}, where $\pi$ is the projection map of the circle fibration over the $D=4$ base $M_{(4)}$.
We refer the readers to \cite{BenettiGenolini:2025icr} for a detailed derivation, simply stating the resulting expression for the $D=5$ on-shell action:
\begin{align}\label{ItotFPbdy}
	\Itot_{(5)}[M_{(5)}] & = I_{(4)}^{\rm FP}[M_{(4)}] - \frac{1}{8\pi G_{(5)}} \int_{\partial M_{(5)}} \KKForm \wedge \FourdSUSYForm \wedge \left( \Phi_2 + \Phi_0 \, \rd \FourdSUSYForm \right) \nn\\
	& - \frac{\Delta x_5}{16\pi G_{(5)}} \int_{M_{(4)}} \Lambda_4 + I_{\mathrm{GHY}}^{\partial M_{(5)}}+ I_{(5)}^{\partial M_{(5)}}\,.
\end{align}
Notice that $\Itot_{(5)}[M_{(5)}]$ is expressed in terms of fixed point contributions $I_{(4)}^{\rm FP}[M_{(4)}]$, and boundary terms. We highlight that $I_{(5)}^{\partial M_{(5)}}$ contain counterterms that remove divergences, as well as finite counterterms, that are in general scheme dependent.

The localised fixed point contribution $I^{\rm FP}_{(4)}[M_{(4)}]$ is a sum over fixed points of $\xi$,
which for $U(1)^3$ gauged supergravity is given by
\begin{equation}\label{I4FP}
	I^{\rm FP}_{(4)}[M_{(4)}] = \frac{\Delta x^5 \pi}{4G_{(5)}} \ii \sum_{\substack{ {\rm fixed} \\ {\rm points} }} \frac{1}{d} \frac{1}{b_1 b_2} \frac{\check{\Phi}_0^1 \check{\Phi}_0^2 \check{\Phi}_0^3}{\Phi_0^0} \,,
\end{equation}
subject to the constraint
\begin{equation}\label{IRconstraint}
	2 Q^{(\ell)} \Phi_0^0 - \check{\Phi}_0^1 - \check{\Phi}_0^2 - \check{\Phi}_0^3 = \kappa (b_1 - \chi b_2) \,.
\end{equation}
Here, $b_1, b_2$ are the weights of $\xi$ at each of its fixed point, and $\kappa, \chi \in \{ \pm 1\}$ are signs, all of which can be determined in a systematic way \cite{BenettiGenolini:2024hyd,BenettiGenolini:2024lbj}.
The weights of $\cK$ also determines $\Phi_0^0$ at the fixed points, extending an argument in \cite{BenettiGenolini:2024kyy}.
We note that $M_{(4)}$ may in general have orbifold singularities, where $d \in \mathbb{N}$ is the order of the orbifold structure group.
$Q^{(\ell)}$ is defined by the charge of the $D=5$ Killing spinor $\zeta$ with respect to $\ell$,
\begin{equation}
	\cL_\ell \zeta = \ii Q^{(\ell)} \zeta \,.
\end{equation}

\subsection{Removing \texorpdfstring{$E_{\rm susy}$}{Esusy} from the partition function}\label{gluing}
We will take $M_{(5)}$ to have the black hole topology $\R^2 \times M_3$, where the $\R^2$ disc factor smoothly caps off at the horizon. Its conformal boundary is given by $S^1_\beta \times M_3$, i.e. the field theory backgrounds we constructed in section \ref{fieldtheory}.

The AdS/CFT correspondence states that, in the appropriate large $N$ limit of the field theory,\footnote{In the Cardy-like limit, the black hole saddle is singled out even at finite $N$ \cite{Cabo-Bizet:2019osg}. Taking the large $N$ limit, we assume that the canonical black hole is the dominant saddle, even away from the Cardy-like limit. Note $\beta$ is kept finite throughout this section.}
\begin{equation}\label{AdSCFT}
	- \log Z_{S^1_\beta \times M_3} = \Itot_{(5)} [M_{(5)}] \,,
\end{equation}
where the supergravity action $\Itot_{(5)} [M_{(5)}]$ is regularised in a supersymmetric scheme.
However, holographic renormalisation in $D=5$ that implements such a scheme involves non-standard, finite counterterms that enter \eqref{ItotFPbdy}, which are not known for general supersymmetric solutions \cite{BenettiGenolini:2016qwm, BenettiGenolini:2016tsn}, including the black holes. Verifying the relation \eqref{AdSCFT} is beyond the scope of the current manuscript.
Instead, we may utilise the following prescription to directly compute the supersymmetric index $\cI_{S^1_\beta \times M_3}$,
\begin{equation}\label{ItotMN}
	- \log \cI_{S^1_\beta \times M_3} = \Itot_{(5)}[M_{(5)}] - \Itot_{(5)}[N_{(5)}] \,,
\end{equation}
where the RHS computes a regularised on-shell action of the background $M_{(5)}$, relative to that of a reference background $N_{(5)}$.
The validity of \eqref{ItotMN} hinges on the following question: what is the correct choice of $N_{(5)}$, such that $\Itot_{(5)}[N_{(5)}]$ precisely accounts for the difference between the partition function and the index?

Note the field theory partition function on these backgrounds are related to the supersymmetric index by \cite{Assel:2014paa,ArabiArdehali:2015iow,ArabiArdehali:2019tdm}
\begin{equation}\label{ZEsusyI}
	\log Z_{S^1_\beta \times M_3} = -\beta E_{\rm susy} + \log \cI_{S^1_\beta \times M_3} \,,
\end{equation}
where $E_{\rm susy}$ is the supersymmetric Casimir energy on $S^1_\beta \times M_3$ \cite{Assel:2015nca}.
The results of \cite{BenettiGenolini:2016tsn} show that, for large classes of $S^1_\beta \times M_3$ backgrounds, there are gravitational fillings with conformal boundary $S^1_\beta \times M_3$, whose supergravity on-shell action precisely agrees with $\beta E_{\rm susy}$.\footnote{The regularised supergravity on-shell action agrees with $\beta E_{\rm susy}$, up to the additional finite counterterms of \cite{BenettiGenolini:2016qwm, BenettiGenolini:2016tsn} (see also \cite{Papadimitriou:2017kzw, An:2017ihs, Papadimitriou:2019gel, Papadimitriou:2019yug, Closset:2019ucb, Kuzenko:2019vvi, Katsianis:2020hzd,Bzowski:2020tue}). Such subtleties may render $E_{\rm susy}$ to be scheme dependent \cite{Panopoulos:2023cdg}, however the relation \eqref{ItotMN} is unchanged.\label{footnote:scanomaly}}
By taking $N_{(5)}$'s to be precisely these backgrounds, we thus expect \eqref{ItotMN} to hold.
The holographic match we show is then between the supersymmetric index on $S^1_\beta \times M_3$, and the on-shell action of a \emph{closed} manifold, obtained by gluing $M_{(5)} = \R^2 \times M_3$ and the $N_{(5)}$'s of \cite{BenettiGenolini:2016tsn} along their common $S^1_\beta \times M_3$ conformal boundary.

For asymptotically AdS black holes, the prescription \eqref{ItotMN} is precisely the ``background subtraction" method utilised in \cite{Colombo:2025ihp,BenettiGenolini:2025icr}.
The black holes of \cite{Gutowski:2004ez,Chong:2005hr} studied in these references have $S^1_\beta \times S^3$ conformal boundary, i.e. the background studied in section \ref{subsection:S1S3}. $N_{(5)}$ in this case corresponds to the AdS$_5$ vacuum.
For asymptotically \emph{locally} AdS black holes, $N_{(5)}$ is no longer global AdS$_5$. We shall refrain from using the term ``background subtraction", as not all AlAdS$_5$ spacetimes can be embedded in an ambient spacetime.
For example, the supersymmetric black holes of \cite{Cassani:2018mlh,Bombini:2019jhp} have biaxially squashed $S^1_\beta \times S^3_v$ conformal boundary. The results of \cite{BenettiGenolini:2016tsn} implies that the choice of $N_{(5)}$, such that \eqref{ItotMN} is valid, is given by a supersymmetric horizonless AlAdS$_5$ solution whose conformal boundary is $S^1_\beta \times S^3_v$.
We highlight that numerical solutions in this class exist in $D=5$ minimal gauged supergravity \cite{Cassani:2014zwa}.
However, this is \emph{not} the solution we will take for $N_{(5)}$. First, the conformal boundary of the solution of \cite{Cassani:2014zwa} does not involve twistings, as highlighted below \eqref{S1S3vmetric}. Second, the non-trivial graviphoton field $\cA$ in \cite{Cassani:2014zwa} has an intrinsically different asymptotic behaviour to what is needed here, as the gauge choice there is such that the Killing spinors are independent of the Euclidean time direction. As emphasised in section \ref{fieldtheory}, our choice of spin structure around $S^1_\beta$ is \emph{anti-periodic}! We demand that the boundary Killing spinors \eqref{S1S3KSchoice}, \eqref{S1S3vKS}, and \eqref{S1S3bKS} on $S^1_\beta \times M_3$ extend to both $M_{(5)}$ and $N_{(5)}$, such that $\cK$ is a Killing vector on the \emph{closed} manifold $M_{(5)} \cup ( - N_{(5)})$.
We can be more explicit by writing the five-dimensional metric on $M_{(5)}$ and $N_{(5)}$ in Fefferman--Graham (FG) form,
\begin{align}\label{FGmetric}
	& {\rd}s^2_{(5)} = g_{\mu\nu}{\rd}y^\mu{\rd}y^\nu = \frac{{\rd}z^2}{z^2} + \frac{1}{z^2} h_{ij}(y,z) {\rd}y^i {\rd}y^j \,, \nn \\
	& h(y,z) = h^{(0)} + z^2 h^{(2)} + z^4 h^{(4)} + \tilde{h}^{(4)} z^4 \log z^2 + \dots
\end{align}
where the conformal boundary is at $z=0$.
The gauge field admits the expansion
\begin{equation}\label{FGgaugefield}
	\cA^I(x,z) = \cA^I_{(0)} + z^2 \cA^I_{(2)} + \tilde{\cA}^I_{(2)} z^2 \log z^2 + \dots
\end{equation}
where by definition $\cA^I_{(0)} = \cA^I_{\rm bdy}$. The boundary metric $h^{(0)}$ can be viewed as a metric on $( M_{(5)} \cup (-N_{(5)}) )\, \backslash \, {\rm Int} \left( \overline{M_{(5)}} \sqcup \overline{N_{(5)}} \right)$.

The name of the game is then to find gravitational fillings $M_{(5)}$ and $N_{(5)}$, for a given $S^1_\beta \times M_3$ background, such that $\Itot_{(5)}[N_{(5)}]$ cancels $\beta E_{\rm susy}$, resulting in the relation \eqref{ItotMN}.
As emphasised in section \ref{fieldtheory}, the Killing spinors on $S^1_\beta \times M_3$ are anti-periodic around $S^1_\beta$, such that they naturally extend to Killing spinors on $M_{(5)} \cup ( - N_{(5)})$.
The field theory backgrounds thus fix a set of boundary conditions, namely the boundary value $\cA^I_{(0)}$ of the bulk gauge field $\cA^I$, the representative $h^{(0)}$ of the conformal structure $[h^{(0)}]$, and finally the boundary value $\epsilon$ of the bulk Killing spinor $\zeta$ (up to a Weyl transformation).
In the remainder of this section, we show that the unintegrated superconformal anomaly vanishes on these backgrounds -- despite complex metrics and anti-periodic Killing spinors -- as expected for a Euclidean supersymmetric background that admits two supercharges of opposite R-charge \cite{Cassani:2013dba}.
Extending arguments in \cite{Papadimitriou:2005ii}, this implies that the boundary conditions yield a well-posed variational problem, for both supergravity fillings $M_{(5)}$ and $N_{(5)}$.

\subsubsection{\texorpdfstring{$\R^2 \times S^3 \, \cup \, (- S^1 \times \R^4)$}{R2×S3 ∪ (−S1×R4)}}
We first consider the supersymmetric black holes of \cite{Gutowski:2004ez,Chong:2005hr}. The horizon is generated by $V_H = \partial_{t_E}$. Their complex, non-extremal deformations are supersymmetric solutions of minimal gauged supergravity with topology $M_{(5)} = \R^2 \times S^3$ \cite{Cabo-Bizet:2018ehj}. Choosing a regular gauge for $\cA^I$, such that $\left. V_H \hook \cA^I \right\vert_{\rm horizon} = 0$ where the $\R^2$ disc factor smoothly caps off, the asymptotics of the solution at leading order as $z \rightarrow 0$ is given by
\begin{align}\label{asympS1S3}
	{\rd}s^2_{(5)} & \sim \frac{{\rd}z^2}{z^2} + \frac{1}{z^2} \Bigg[ \frac{\beta^2}{4\pi^2} {\rd} t_E^2 + {\rd} \vartheta^2 \nn \\
	& \qquad \qquad \qquad + \sin^2 \vartheta \left({\rd} \varphi_1 - \frac{\ii \beta}{2\pi} \Omega_1 {\rd} t_E \right)^2 + \cos^2 \vartheta \left({\rd}\varphi_2 - \frac{\ii \beta}{2\pi} \Omega_2 {\rd} t_E \right)^2 \Bigg] \,, \nn \\
	\cA^R_{(0)} & = \frac{\ii \beta}{2\pi} \Psi \, {\rd}t_E \,,
\end{align}
with
\begin{equation}\label{asympS1S3cond}
	\frac{\beta}{2\pi\ii} ( \Omega_1 + \Omega_2 - \Psi_1 - \Psi_2 - \Psi_3 +1) = c_R \,, \qquad c_R = \pm 1 \,.
\end{equation}
Note we rescaled the coordinate $t_E$, such that $t_E \sim t_E + 2\pi$.
The sign $c_R$ is associated to the two ``branches" of the black hole solutions in the nomenclature of \cite{Aharony:2021zkr}, where in \cite{BenettiGenolini:2025icr} the discrete data was identified with the charge of the $D=5$ Killing spinor $\zeta$ with respect to $V_H$.
Choosing $n = - c_R$ in \eqref{KSE_4d_constraint_S3}, the Killing spinor $\zeta$ of the black hole is precisely the bulk extension of the boundary (conformal) Killing spinor \eqref{S1S3KSchoice}. 
Notice that the usual global AdS$_5$ background with \emph{periodic} Killing spinors around the Euclidean time circle does \emph{not} satisfy the boundary conditions specified by \eqref{asympS1S3}-\eqref{asympS1S3cond}, and hence cannot be $N_{(5)}$. Indeed, the regularised black hole on-shell action used in \cite{Cabo-Bizet:2018ehj} was obtained by ``AdS subtraction" \cite{Chen:2005zj}, where the metric is obtained by setting the mass and charge to zero in the black hole metric. In the supersymmetric limit, $N_{(5)}$ obtained this way has topology $S^1_\beta \times \R^4$, and admits Killing spinors that are \emph{anti-periodic} around $S^1_\beta$.\footnote{To perform the gluing, one is required to match the period of the Euclidean time coordinate of the black hole and AdS metrics. A similar idea was recently used in \cite{Kim:2025ziz}.}

By writing \eqref{asympS1S3} we are only fixing the conformal structure, and not the representative \cite{Papadimitriou:2005ii}. The boundary gauge field is flat, and from the boundary metric $h^{(0)}$ we can easily check that
\begin{equation}\label{anom_S1S3}
	E = C_{ijkl} C^{ijkl} = 0 \,,
\end{equation}
where $E$ is the Euler density, and $C_{ijkl}C^{ijkl}$ is the square of the Weyl tensor, given by
\begin{align}
	E & = R_{ijkl} R^{ijkl} - 4 R_{ij} R^{ij} + R^2 \,, \nn \\
	C_{ijkl}C^{ijkl} & = R_{ijkl} R^{ijkl} - 2 R_{ij} R^{ij} + \frac{1}{3} R^2 \,.
\end{align}

\subsubsection{\texorpdfstring{$\R^2 \times S^3_v \, \cup \, (- S^1 \times \R^4_v)$}{R2×S3v ∪ (−S1×R4v)}}
Next, we consider the supersymmetric black holes of \cite{Cassani:2018mlh,Bombini:2019jhp}, with biaxially squashed $S^1_\beta \times S^3_v$ boundary. While complex, non-extremal deformations of these solutions have not been constructed, the physics of such Euclidean saddles have been explored restricting to equal angular momenta $\Omega_1 = \Omega_2$, based on an $SU(2) \times U(1)$ invariant ansatz \cite{Ntokos:2021duk}. Explicit bulk Killing spinors have been constructed in \cite[app. C]{Ntokos:2021duk}. Here, we assume that the supersymmetric non-extremal solutions exist, even for $\Omega_1 \neq \Omega_2$. We shall refer to these as the $\R^2 \times S^3_v$ saddles.

With regularity conditions imposed, the black hole solutions should have the following asymptotic behaviour in the leading order as $z \rightarrow 0$, 
\begin{align}\label{asympS1S3v}
	{\rd}s^2_{(5)} & \sim \frac{{\rd}z^2}{z^2} + \frac{1}{z^2} \Bigg[ \frac{\beta^2}{4\pi^2} {\rd} t_E^2 + {\rd} \vartheta^2 \nn \\
	& \qquad \qquad \qquad + \sin^2 \vartheta \left({\rd} \varphi_1 - \frac{\ii \beta}{2\pi} \Omega_1 {\rd} t_E \right)^2 + \cos^2 \vartheta \left({\rd}\varphi_2 - \frac{\ii \beta}{2\pi} \Omega_2 {\rd} t_E \right)^2 \nn \\
	& \qquad \qquad \qquad + (v^2-1) \left( \sin^2\vartheta ({\rd}\varphi_1 - \frac{\ii \beta}{2\pi} \Omega_1 {\rd}t_E) + \cos^2\vartheta ({\rd}\varphi_2 - \frac{\ii \beta}{2\pi} \Omega_2 {\rd}t_E) \right)^2 \Bigg] \,, \nn \\
	\cA^R_{(0)} & =  \frac{\ii \beta}{4\pi} \left[ \Omega_1 + \Omega_2 + v - \left( 3 - \frac{2}{v} - 2\Psi +  \Omega_1 + \Omega_2 \right) \right] {\rd} t_E \nn \\
	& + (v^2 - 1) \Big[ \frac{\ii\beta}{2\pi} (\Omega_1 \sin^2\vartheta + \Omega_2 \cos^2\vartheta) {\rd} t_E - \sin^2\vartheta {\rd}\varphi_1 - \cos^2\vartheta {\rd}\varphi_2 \Big]  \,,
\end{align}
where
\begin{equation}\label{asympS1S3vcond}
	\frac{\beta}{2\pi\ii} \left( \Omega_1 + \Omega_2 - \frac{2}{v} - \Psi_1 - \Psi_2 - \Psi_3 + 3 \right)  = c_R \,,\qquad c_R = \pm 1 \,.
\end{equation}
The metric \eqref{asympS1S3v} is asymptotically locally AdS, and a straightforward calculation shows that these solve the supergravity equations of motion at leading order as $z \rightarrow 0$.
From $h^{(0)}$, one can check that $E=0$ is unchanged from \eqref{anom_S1S3}.
The superconformal anomaly vanishes, though in a non-trivial way:
\begin{equation}
	C_{ijkl} C^{ijkl} = \frac{8}{3} \cF^{(0)}_{ij} \cF^{(0)ij} = \frac{64}{3} (v^2 - 1)^2 \, .
\end{equation}

\subsubsection{\texorpdfstring{$\R^2 \times S^3_{\mf{b}_1,\mf{b}_2} \, \cup \, (- S^1 \times \R^4_{\mf{b}_1,\mf{b}_2})$}{R2×S3b1b2 ∪ (−S1×R4b1b2)}}
Finally, we consider complex, supersymmetric, non-extremal black holes whose conformal boundary is given by the $S^1_\beta \times S^3_{\mf{b}_1,\mf{b}_2}$ background constructed in section \ref{subsection:S1M3}. We shall refer to these as the $\R^2 \times S^3_{\mf{b}_1,\mf{b}_2}$ saddles.
For $N_{(5)}$, we assume that there exist supersymmetric AlAdS$_5$ solutions with the same conformal boundary, with $S^1_\beta \times \R^4$ topology, that admit Killing spinors that are anti-periodic around $S^1_\beta$.
Solutions for both $M_{(5)}$ and $N_{(5)}$ are required to take the following form, at leading order as $z \rightarrow 0$:
\begin{align}\label{asympS1S3b}
	{\rd}s^2_{(5)} & \sim \frac{{\rd}z^2}{z^2} + \frac{1}{z^2} \Bigg[ \frac{\beta^2}{4\pi^2} {\rd} t_E^2 + f(\vartheta)^2 {\rd} \vartheta^2 \nn \\
	& \qquad \qquad \qquad + \frac{\sin^2 \vartheta}{\mf{b}_1^2} \left({\rd} \varphi_1 - \frac{\ii \beta}{2\pi} \Omega_1 {\rd} t_E \right)^2 + \frac{\cos^2 \vartheta}{\mf{b}_2^2} \left({\rd}\varphi_2 - \frac{\ii \beta}{2\pi} \Omega_2 {\rd} t_E \right)^2 \Bigg] \,, \nn \\
	\cA^R_{(0)} & = \frac{\ii\beta}{4\pi} \left[ \frac{1}{f(\vartheta)} \left( \frac{\Omega_1}{\mf{b}_1} + \frac{\Omega_2}{\mf{b}_2} +1  \right) - \left( 3 - \mf{b}_1 - \mf{b}_2 - 2\Psi +  \Omega_1 + \Omega_2 \right) \right] {\rd} t_E \nn \\
	& + \frac{1}{2} \left( 1- \frac{1}{f(\vartheta) \mf{b}_1}\right) {\rd}\varphi_1 + \frac{1}{2} \left( 1- \frac{1}{f(\vartheta) \mf{b}_2}\right) {\rd}\varphi_2 \,,
\end{align}
where
\begin{equation}
	\frac{\beta}{2\pi\ii} \left( \Omega_1 + \Omega_2 - \mf{b}_1 - \mf{b}_2 - \Psi_1 - \Psi_2 - \Psi_3 + 3 \right)  = c_R \,,\qquad c_R = \pm 1 \,.
\end{equation}
The five-dimensional metric \eqref{asympS1S3b} is asymptotically locally AdS, and a straightforward calculation shows that these solve the supergravity equations of motion at leading order as $z \rightarrow 0$. This, of course, is not sufficient to guarantee that such supersymmetric AlAdS$_5$ solutions exist. While the sub-leading term $h^{(2)}$ in the FG expansion is fixed by $h^{(0)}$ \cite{deHaro:2000vlm}, one needs to verify the existence of supersymmetric solutions $M_{(5)} = \R^2 \times S^3_{\mf{b}_1,\mf{b}_2}$ and $N_{(5)} = S^1_\beta \times \R^4_{\mf{b}_1,\mf{b}_2}$, respectively at all orders. Equivariant localization is agnostic of such challenges; we will find that the gravity result precisely matches the elliptically squashed index \eqref{HondaSquashedIndexb} in the large $N$ limit, both of which are evaluated solely using boundary data.

The squashing does not affect the Euler density, and one can check that we have $E=0$ as expected from \eqref{anom_S1S3}. From $h^{(0)}$ and $\cA^{(0)}$, we can show that the superconformal anomaly vanishes, again in a non-trivial way:
\begin{equation}
	C_{ijkl} C^{ijkl} = \frac{8}{3} \cF^{(0)}_{ij} \cF^{(0)ij} = \frac{4 \left( \tan^2\vartheta + \csc^2\vartheta \right)f'(\vartheta)^2}{3 f(\vartheta)^6} \, .
\end{equation}

\subsection{SUSY index from fixed point contributions}\label{subsection:indexfromFP}

The gluing of $M_{(5)}$ and $N_{(5)}$ has desirable properties that further simplify \eqref{ItotMN}.
Recall from \eqref{ItotFPbdy} that $\Itot_{(5)}$ \emph{a priori} includes boundary terms, in addition to fixed point contributions \eqref{I4FP}. It is only the latter that can be computed without solving any supergravity equations, using equivariant localization. However, because we glued the two geometries along their common conformal boundary, the boundary terms that depend only on the intrinsic geometry automatically cancel.\footnote{Note this leaves us with fixed point contributions, and an integral of $\Lambda_4$ over $M_{(4)} \cup (-N_{(4)})$.
The latter cancels, assuming $\cA^I$ are in a regular gauge on $M_{(5)}$. On $N_{(5)}$ the gauge fields are globally defined by construction, hence $\Lambda_4$ is exact on both $M_{(4)}$ and $N_{(4)}$.}
Hence, we conclude that \eqref{ItotMN} reduces to
\begin{equation}\label{indexfromFP}
	-  \log \cI_{S^1_\beta \times M_3} = I_{(4)}^{\rm FP} [M_{(4)}] - I_{(4)}^{\rm FP} [N_{(4)}] \,.
\end{equation}

As explained in the last subsection, we will assume that the boundary Killing spinor bilinears constructed on the backgrounds $S^1_\beta \times S^3$, $S^1_\beta \times S^3_{\mf{b}_1,\mf{b}_2}$, and $S^1_\beta \times S^3_v$ extend to the respective supersymmetric Killing vectors $\cK$ on the double-sided gravitational fillings $M_{(5)} \cup (- N_{(5)})$.
Noting that we rescaled the coordinate $t_E$ such that $t_E \sim t_E + 2\pi$, from \eqref{S1S3susyKV}, \eqref{S1S3vsusyKV}, and \eqref{S1S3bsusyKV} we have
\begin{align}
	\cK & = \partial_{t_E} - \frac{\beta (\Omega_1 -1)}{2\pi\ii} \partial_{\varphi_1} - \frac{\beta (\Omega_2 -1)}{2\pi\ii} \partial_{\varphi_2} \,, \nn \\
	& = \partial_{t_E} - \sigma \partial_{\varphi_1} - \tau \partial_{\varphi_2} \,, 
\end{align}
on $\R^2 \times S^3 \, \cup \, (- S^1 \times \R^4)$,
\begin{align}
	\cK & = \partial_{t_E} - \frac{\beta}{2\pi\ii} \left( \Omega_1 - \frac{1}{v} \right) \partial_{\varphi_1} - \frac{\beta}{2\pi\ii} \left(\Omega_2 - \frac{1}{v} \right) \partial_{\varphi_2} \,, \nn \\
	& = \partial_{t_E} - \sigma_v \partial_{\varphi_1} - \tau_v \partial_{\varphi_2} \,, 
\end{align}
on $\R^2 \times S^3_v \, \cup \, (- S^1 \times \R^4_v)$,
and
\begin{align}
	\cK & = \partial_{t_E} - \frac{\beta (\Omega_1 -\mf{b}_1)}{2\pi\ii} \partial_{\varphi_1} - \frac{\beta (\Omega_2 - \mf{b}_2)}{2\pi\ii} \partial_{\varphi_2} \,, \nn \\
	& = \partial_{t_E} - \sigma_{\mf{b}} \partial_{\varphi_1} - \tau_{\mf{b}} \partial_{\varphi_2} \,, 
\end{align}
on $\R^2 \times S^3_{\mf{b}_1,\mf{b}_2} \, \cup \, (- S^1 \times \R^4_{\mf{b}_1,\mf{b}_2})$,
with complex chemical potentials \eqref{susychempot_S3}, \eqref{susychempot_S3v}, and \eqref{susychempot_S3b}, respectively.
For convenience, we will collectively denote $\cK$ as
\begin{equation}\label{susycK}
	\cK = \partial_{t_E} + \varepsilon_1 \partial_{\varphi_1} + \varepsilon_2 \partial_{\varphi_2} \,,
\end{equation}
express the fixed point contributions \eqref{I4FP} in terms of $(\varepsilon_1, \varepsilon_2)$, and identify
\begin{equation}\label{weightsid}
\begin{split}
	& \R^2 \times S^3 \, \cup \, (- S^1 \times \R^4) : \qquad (\varepsilon_1, \varepsilon_2) \leftrightarrow - (\sigma, \tau) \,, \\
	& \R^2 \times S^3_v \, \cup \, (- S^1 \times \R^4_v) : \qquad (\varepsilon_1, \varepsilon_2) \leftrightarrow - (\sigma_v, \tau_v) \,, \\
	& \R^2 \times S^3_{\mf{b}_1,\mf{b}_2} \, \cup \, (- S^1 \times \R^4_{\mf{b}_1,\mf{b}_2}) : \qquad (\varepsilon_1, \varepsilon_2) \leftrightarrow - (\sigma_{\mf{b}}, \tau_{\mf{b}}) \,,
\end{split}
\end{equation}
at the end of all computations.

To proceed, we consider a generic KK reduction along
\begin{equation}\label{ptwist}
	\ell = p \, \partial_{t_E} + \left( \partial_{\varphi_1} + \partial_{\varphi_2} \right) \,,
\end{equation}
where $\Delta x^5 = 2\pi$, and $p \in \Z$ is an arbitrary integer.
Given $M_{(5)} \cong \R^2 \times S^3$ and $N_{(5)} \cong S^1 \times \R^4$,\footnote{The symbol $\cong$ means ``diffeomorphic to", i.e. $M_3 \cong S^3$ collectively refers to the round $S^3$, as well as the squashed ones.}
the base space of the circle fibrations are, respectively,
\begin{equation}
	M_{(4)} \cong \cO(-p) \rightarrow S^2 \,, \qquad N_{(4)} \cong \R^4 / \Z_p \,.
\end{equation}
We emphasise that these are topological data, and hence are insensitive to the squashings.
From the anti-periodic spin structure of the $\R^2$ disc factor, we have
\begin{equation}\label{Lietau}
	\cL_{\partial_{t_E}} \zeta = c_R \frac{\ii}{2} \zeta \,, \qquad c_R = \pm 1\,,
\end{equation}
and from \eqref{app:LieKS}, \eqref{app:LieKSb}, and \eqref{app:LieKSv}, we have
\begin{equation}\label{LieJ}
	\cL_{\partial_{\varphi_1}} \zeta = \cL_{\partial_{\varphi_2}} \zeta = c_J \frac{\ii}{2} \zeta \,, \qquad c_J = \pm 1\,.
\end{equation}
The spinor charge $Q^{(\ell)}$ is then given by
\begin{equation}
	Q^{(\ell)} = \frac{p}{2} c_R + c_J \,.
\end{equation}

The remaining data that enters \eqref{indexfromFP} via \eqref{I4FP} are the signs $\kappa, \chi \in \{ \pm 1 \}$ and the weights $b_1, b_2, \Phi_0^0$ at the fixed points.
These can be systematically computed as explained in \cite{BenettiGenolini:2024hyd}.
The weights explicitly depend on the squashings, however the functional dependence when expressed in terms of the variables $(\varepsilon_1, \varepsilon_2)$ are identical, keeping in mind the identifications \eqref{weightsid}.
Hence we simply state the results, and refer the readers to \cite{BenettiGenolini:2025icr} for a detailed derivation, noting that $\vartheta^{\rm here} = \vartheta^{\rm there} + \pi/2$, which effectively exchanges the ``north" and ``south" poles.

On $M_{(4)}$, the Killing vector $\xi = \pi_*(\cK)$ has fixed points at the north and south poles of the zero-section $S^2 \subset \cO(-p) \rightarrow S^2$.
The fixed point data is given by \cite{BenettiGenolini:2025icr}
\begin{equation}
\label{eq:BH_weights}
\begin{split}
	& (b_1^N, b_2^N) = ( \varepsilon_1 - \varepsilon_2 , -1 + \varepsilon_2 p) \, , \\
	& (b_1^S , b_2^S) = ( -1 + \varepsilon_1 p , - \varepsilon_1 + \varepsilon_2 ) \,, \\
	& \left.\Phi_0^0\right\vert_N = \varepsilon_2 \,, \qquad \left.\Phi_0^0\right\vert_S = \varepsilon_1 \,.
\end{split}
\end{equation}
On $N_{(4)} \cong \R^4/\Z_p$, there is a single fixed point at the centre. We have
\begin{equation}
\begin{split}
	& b_1^o = - \frac{1}{p} \det(\vec{v}_1,\xi) = -\frac{1}{p} + \varepsilon_1 \,, \\
	& b_2^o = \frac{1}{p} \det(\vec{v}_0,\xi) = -\frac{1}{p} + \varepsilon_2 \,, \\
	& \left.\Phi_0^0\right\vert_o = \frac{1}{p} \,.
\end{split}
\end{equation}
Finally, our field theory conventions are such that $c_R = -1$ and $c_J = +1$. This fixes the signs (cf. \cite[(4.24)]{BenettiGenolini:2025icr})
\begin{equation}
	\chi_N = \chi_S = c_R c_J = -1 \,, \qquad \kappa_N = - c_J = -1 \,, \qquad \kappa_S = c_R = -1 \,,
\end{equation}
as well as $\kappa_o = - c_J = +1$, $ \chi_o = -1$.

Collecting everything, the constraint \eqref{IRconstraint} reduces to
\begin{align}
	\varepsilon_1 + \varepsilon_2 - \left. \check{\Phi}_0^1 \right\vert_{\rm IR} - \left. \check{\Phi}_0^2 \right\vert_{\rm IR} - \left. \check{\Phi}_0^3 \right\vert_{\rm IR} = 1 \,, 
\end{align}
at all three fixed points, where we defined $\left.\check{\Phi}_0^I\right\vert_{\rm IR} \equiv \left.\check{\Phi}_0^I\right\vert_N=\left.\check{\Phi}_0^I\right\vert_S=\left.\check{\Phi}_0^I\right\vert_o$, extending an argument in \cite{BenettiGenolini:2025icr}.
Subject to this constraint, the fixed point contributions \eqref{I4FP} evaluate to
\begin{align}
	I_{(4)}^{\rm FP}[M_{(4)}] & = I_{(4)N}^{\rm FP} + I_{(4)S}^{\rm FP} \nn \\
	& = - \frac{\ii \pi^2}{2 G_{(5)}} \left[ \frac{ \left. \check{\Phi}_0^1 \check{\Phi}_0^2 \check{\Phi}_0^3 \right\vert_N }{\varepsilon_2 (\varepsilon_1 - \varepsilon_2) (1-p\varepsilon_2)} - \frac{ \left. \check{\Phi}_0^1 \check{\Phi}_0^2 \check{\Phi}_0^3 \right\vert_S }{\varepsilon_1 (\varepsilon_1 - \varepsilon_2) (1-p\varepsilon_1)} \right] \,,
\end{align}
on $M_{(4)}$, and
\begin{equation}
	I_{(4)}^{\rm FP}[N_{(4)}] = \frac{\ii \pi^2}{2 G_{(5)}} \frac{\left. \check{\Phi}_0^1 \check{\Phi}_0^2 \check{\Phi}_0^3 \right\vert_o}{(1-p\varepsilon_1)(1-p\varepsilon_2)} p^2 \,,
\end{equation}
on $N_{(4)}$.
From \eqref{indexfromFP}, the supersymmetric index is then given by
\begin{equation}
	\log \cI_{S^1_\beta \times M_3} = \ii \pi N^2 \frac{\left. \check{\Phi}_0^1 \check{\Phi}_0^2 \check{\Phi}_0^3 \right\vert_{\rm IR}}{\varepsilon_1 \varepsilon_2} \,.
\end{equation}
where we have used the AdS/CFT dictionary
\begin{equation}
	\frac{\pi}{2G_{(5)}} = N^2 \,.
\end{equation}
Substituting in the identifications \eqref{weightsid},
we have
\begin{equation}
	\log \cI_{S^1_\beta \times S^3} = \ii \pi N^2 \frac{\left. \check{\Phi}_0^1 \check{\Phi}_0^2 \check{\Phi}_0^3 \right\vert_{\rm IR}}{\sigma \tau} \,, \qquad \sigma + \tau + \left. \check{\Phi}_0^1 \right\vert_{\rm IR} + \left. \check{\Phi}_0^2 \right\vert_{\rm IR} + \left. \check{\Phi}_0^3 \right\vert_{\rm IR} = -1 \,,
\end{equation}
on $\R^2 \times S^3 \, \cup \, (- S^1 \times \R^4)$,
\begin{equation}\label{I_OS_v}
	\log \cI_{S^1_\beta \times S^3_v} =\ii \pi N^2 \frac{\left. \check{\Phi}_0^1 \check{\Phi}_0^2 \check{\Phi}_0^3 \right\vert_{\rm IR}}{\sigma_v \tau_v} \,, \qquad \sigma_v + \tau_v + \left. \check{\Phi}_0^1 \right\vert_{\rm IR} + \left. \check{\Phi}_0^2 \right\vert_{\rm IR} + \left. \check{\Phi}_0^3 \right\vert_{\rm IR} = -1 \,,
\end{equation}
on $\R^2 \times S^3_v \, \cup \, (- S^1 \times \R^4_v)$,
and
\begin{equation}
	\log \cI_{S^1_\beta \times S^3_{\mf{b}_1,\mf{b}_2}} =\ii \pi N^2 \frac{\left. \check{\Phi}_0^1 \check{\Phi}_0^2 \check{\Phi}_0^3 \right\vert_{\rm IR}}{\sigma_{\mf{b}} \tau_{\mf{b}}} \,, \qquad \sigma_{\mf{b}} + \tau_{\mf{b}} + \left. \check{\Phi}_0^1 \right\vert_{\rm IR} + \left. \check{\Phi}_0^2 \right\vert_{\rm IR} + \left. \check{\Phi}_0^3 \right\vert_{\rm IR} = -1 \,,
\end{equation}
on $\R^2 \times S^3_{\mf{b}_1,\mf{b}_2} \, \cup \, (- S^1 \times \R^4_{\mf{b}_1,\mf{b}_2})$.
In the large $N$ limit, these precisely match the field theory results \eqref{roundSCI}, \eqref{HondaSquashedIndex}, and \eqref{HondaSquashedIndexb}, respectively, upon identifying the fixed point variables $\left.\check{\Phi}_0^I\right\vert_{\rm IR}$ with the background holonomies of boundary gauge fields.

Notice the dependence on the KK vector $\ell$, parametrised by the arbitrary integer $p$, dropped out of the final answer. While $I_{(4)}^{\rm FP}[M_{(4)}]$ and $I_{(4)}^{\rm FP}[N_{(4)}]$ individually depend on $\ell$, their difference does not, and corresponds to a physical observable: the supersymmetric index on $S^1_\beta \times M_3$.
More generally, it is possible to start with \emph{any} combination of the $U(1)^3$ isometry, and show that the final answer is $\ell$-independent \cite{Colombo:2025ihp}. For example, setting $p=0$, the reduction \eqref{ptwist} is simply along the Hopf fibre of the $S^3$, and in \cite{BenettiGenolini:2025icr} it was shown that this reduction can be replaced by a countably infinite number of ``spindle reductions", labelled by coprime integers $(n_N, n_S)$, such that $S^1 \hooklongrightarrow S^3 \overset{\pi}{\longrightarrow} \mathbb{WCP}^1_{[n_N,n_S]}$.
The orbifold singularity at the two poles of the spindle dress up various fixed point formulae, however the dependence on $(n_N,n_S)$ drops out of the final answer, as shown in \cite[sect. 4.2]{BenettiGenolini:2025icr}.
The statement generalises for Lens space black holes with $M_{(5)} \cong \R^2 \times L(\tp,\tq)$ and $N_{(5)} \cong S^1 \times \R^4/\Z_\tp$. The circle action generated by $n_N \partial_{\varphi_1} + n_S \partial_{\varphi_2}$ commutes with the $\Z_\tp$ action \eqref{Lpqorbiid}, and thus defines a Seifert fibration. According to Theorem 4.10 of \cite{geiges2017seifertfibrationslensspaces}, the base space of the fibration is generally a ``non-coprime" spindle, in the language of \cite{Arav:2025jee}. The final answer \eqref{indexfromFP} should then again be independent of the choice of fibration, with an overall factor of $1/\tp$ produced from $\Delta x^5$ in \eqref{I4FP}, in agreement with \eqref{HondaLensIndex}.

\section{Discussion}\label{discussion}

In this paper, we studied rigid supersymmetric backgrounds $S^1_\beta \times M_3$, comprising elliptically/biaxially squashed three-spheres and Lens spaces, that admit Killing spinors that are anti-periodic around the Euclidean time circle. Such backgrounds should arise from the conformal boundary of complex, supersymmetric, non-extremal Euclidean black holes, with topology $\R^2 \times M_3$.
We explicitly constructed non-trivial Killing spinors on these backgrounds, together with a choice of profile for the new minimal background fields. Using a 3d effective field theory approach similar to \cite{Cassani:2021fyv,ArabiArdehali:2021nsx}, we computed the supersymmetric index of the $SU(N)$ $\cN=4$ SYM on these backgrounds in a Cardy-like limit \cite{Honda:2019cio}. Moreover, we showed that the superconformal anomaly vanishes on these backgrounds, consistent with expectations for Euclidean supersymmetric backgrounds admitting two supercharges of opposite R-charge.
Assuming supergravity fillings of these backgrounds exist, we demonstrated how the techniques of equivariant localization can be used to precisely recover the field theory results from a gravity computation, without solving any supergravity equations.
Extending the ``background subtraction" method utilised in \cite{Cabo-Bizet:2018ehj,BenettiGenolini:2025icr,Colombo:2025ihp} for asymptotically AdS$_5$ black holes, we proposed the figure \ref{figure:MglueN} prescription for general AlAdS$_5$ black holes, to compute directly the supersymmetric index, effectively removing the contribution from the supersymmetric Casimir energy from the partition function.

We note that the sub-leading correction in the large $N$ limit of the superconformal index \cite{Cassani:2021fyv} has been recovered from four-derivative corrections to the $D=5$ supergravity on-shell action \cite{Bobev:2022bjm,Cassani:2022lrk,Cassani:2024tvk}.
Restricting to $\cN=4$ SYM, we computed the supersymmetric index in the Cardy-like limit on our complex $S^1_\beta \times S^3_v$ and $S^1_\beta \times S^3_{\mf{b}_1,\mf{b}_2}$ backgrounds, which is a finite $N$ result. However, four-derivative corrections in supergravity trivially vanishes for $\cN = 4$ SYM, since ${\rm Tr} R=0$ \cite{Bobev:2022bjm}.
It would be thus interesting to revisit the computations in \cite{Cassani:2021fyv,Ardehali:2021irq,Ohmori:2021dzb}, using the backgrounds constructed in this paper, namely \eqref{S1S3bmetric} \& \eqref{AnmVnm_S1S3b}, as well as \eqref{S1S3vmetric} \& \eqref{AnmVnm_S1S3v}, generalising our results to 4d $\cN=1$ SCFTs for which ${\rm Tr} R$ scales like $N$. We expect the results to take the same form as in \cite[(6.36)]{Cassani:2021fyv}, where the complex structure parameters are replaced by \eqref{susychempot_S3v} and \eqref{susychempot_S3b}, respectively. It will be interesting to see if such results can be recovered by incorporating higher derivatives in equivariant localization.

It will also be interesting to extend the figure \ref{figure:MglueN} prescription to AdS$_7$/CFT$_6$. Using holographic renormalisation, the authors of \cite{Bobev:2025xan} computed the Euclidean on-shell action of the BDHM black hole\cite{Bobev:2023bxl}, and compared their result with the superconformal index of the 6d $\cN=(2,0)$ theory on $S^1 \times S^5$. The on-shell action of the BDHM black hole regularised this way satisfies a quantum statistical relation \cite[(3.29)]{Bobev:2025xan}:
\begin{equation}\label{QSR_BDDH}
	I_{\rm reg} = - S + \beta \left[ E - \sum_{i=1}^3 \Omega_i J_i - \sum_{I=1}^2 \Phi_I Q_I \right] \,.
\end{equation}
If a particular choice is made for the finite counterterms, such that the energy $E = E^{\rm BDHM}$ is zero for global AdS$_7$, $I_{\rm reg} = I_{\rm reg}^{\rm BDHM}$ agrees with minus the logarithm of the index, $\cI_{S^1 \times S^5}$ \cite{Bobev:2025xan}. Subtleties in such counterterms are 6d analogues of the scheme dependence of the partition function of 4d $\cN=1$ SCFTs (recall our footnote \ref{footnote:scanomaly}). In terms of \eqref{QSR_BDDH}, there is a scheme dependence in the on-shell action $I_{\rm reg}$, which compensates that of $E$, while $(S, J_i, Q_I)$ are scheme independent \cite{Papadimitriou:2017kzw}. Indeed, there is another scheme, for which the energy of AdS$_7$ is given by the supersymmetric Casimir energy $E_{\rm susy}$. Replacing $I_{\rm reg}^{\rm BDHM}$ with $(- \log \cI_{S^1 \times S^5})$, \eqref{QSR_BDDH} can be written as
\begin{equation}\label{7dclosed}
	- \log Z_{S^1 \times S^5} = - S + \beta \left[ \left( E^{\rm BDHM} + E_{\rm susy} \right) - \sum_{i=1}^3 \Omega_i J_i - \sum_{I=1}^2 \Phi_I Q_I \right] \,,
\end{equation}
where $\log Z_{S^1 \times S^5} = - \beta E_{\rm susy} + \log \cI_{S^1 \times S^5}$ \cite{Bobev:2015kza}.
The holographic setup we proposed in section \ref{gluing} suggests that $I_{\rm reg}^{\rm BDHM} = - \log \cI_{S^1 \times S^5}$ is the on-shell action of a seven-dimensional closed manifold, obtained by gluing the BDHM black hole with global AdS$_7$ along their common $S^1 \times S^5$ conformal boundary (analogously to figure \ref{figure:MglueN}).
The statements should also apply for general classes of AlAdS$_7$ black holes, including the $L^{p,q,r}$ black holes studied in \cite{Bobev:2025xan}. We expect the field theory analysis presented in this paper to generalise to supersymmetric $S^1_\beta \times M_5$ backgrounds, realising Killing spinors that are anti-periodic around $S^1_\beta$. We hope to return to this problem in future work.

Finally, in $D=5$ gauged supergravity there exist uniqueness theorems \cite{Lucietti:2021bbh,Lucietti:2022fqj,Lucietti:2023mvj} for (asymptotically locally) AdS$_5$ black holes, which state that any supersymmetric toric solution that is timelike outside a \emph{smooth} horizon with compact cross sections is locally isometric to the known black hole of \cite{Gutowski:2004ez,Chong:2005hr} (or its near horizon geometry). The construction of these theorems are based on the Lorentzian classification of \cite{Gauntlett:2003fk}, and hence are not applicable for the complex Euclidean saddles studied in this paper. It would be interesting if the uniqueness results can be extended to cover such solutions. It is worth emphasising that the existence of Lorentzian BPS black hole solutions numerically constructed in \cite{Cassani:2018mlh,Bombini:2019jhp}, with conformal boundary comprising a biaxially squashed three-sphere, are not counterexamples to these uniqueness theorems. It was shown in \cite{Lucietti:2021bbh} that the squashed black holes have \emph{non-smooth} horizons: they are $C^1$, but not $C^2$. Whether supersymmetric black hole solutions with $\Omega_1 \neq \Omega_2$ and biaxially squashed $S^1 \times S^3_v$ conformal boundary exist, or whether any supersymmetric black hole solution with elliptically squashed $S^1 \times S^3_{\mf{b}_1, \mf{b}_2}$ conformal boundary exists, remain as an open problem. The uniqueness results imply that if the Lorentzian solutions were to exist, they will also have non-smooth horizons. Complex, non-extremal deformations of these solutions will \emph{not} have this feature \cite{Horowitz:2022mly}, and should be taken seriously as dominant saddles of the Euclidean gravitational path integral \cite{BenettiGenolini:2025jwe}, given the holographic match with supersymmetric indices as presented in this paper.
The matching persists in the BPS limit, where the complex Euclidean saddles can be Wick-rotated to real Lorentzian black holes. While squashing the boundaries of supersymmetric AdS black holes result in singular horizons, holography seems to be suggesting that the black holes are just as physical.

\section*{Acknowledgements}

\noindent
I am grateful to Masazumi Honda, Yusheng Jiao, Seok Kim, Rishi Mouland, Jesse van Muiden, Sameer Murthy, Pantelis Panopoulos, Ioannis Papadimitriou, and James Sparks for helpful discussions.
Special thanks goes to Pietro Benetti Genolini and Jerome Gauntlett for inspiring discussions and valuable comments on a draft. I am also grateful to Davide Cassani, Vasil Dimitrov, and Luigi Tizzano for comments on an earlier version of this paper.

I would like to thank the Royal Society under the International Collaboration Award Grant {\textbackslash}R2{\textbackslash}242058 and the RIKEN Interdisciplinary Theoretical and Mathematical Sciences Program. Discussions during the ``Japan-UK Workshop on Quantum Gravity" were useful in completing this work.
I am supported by a Dean's PhD studentship at Imperial College.

\appendix

\section{Squashed partition function and the index}\label{app:secondsheet}
In section \ref{fieldtheory}, we utilised Honda's formula \eqref{HondaIndex} to compute the supersymmetric index on complex backgrounds $S^1_\beta \times S^3_v$ and $S^1_\beta \times S^3_{\mf{b}_1, \mf{b}_2}$, having worked out a choice of profile for background fields that admit Killing spinors that are anti-periodic around the Euclidean time circle. An alternative approach is to start from a real background with periodic Killing spinors, and realise the constraint on the chemical potentials by going to the ``second sheet" \cite{Cassani:2021fyv}. Here, we briefly comment on the relation between the two approaches.

The careful reader may worry that the ``squashed" backgrounds do not preserve the full $\cN=1$ superconformal algebra. Recall that, even when the $S^3$ is \emph{round}, it is only a subalgebra that matters for the superconformal index; one can pick a complex supercharge obeying the following algebra \cite{Romelsberger:2005eg,Kinney:2005ej}
\begin{equation}\label{QQbar}
	\{ \cQ, \overline{\cQ} \} = H - J_1 - J_2 - \frac{3}{2} R \,,
\end{equation}
and define a (refined) Witten index of the specific supercharge $\cQ$ by the trace
\begin{equation}\label{SCIwrittenout}
	\cI(\sigma_{\rm FT}, \tau_{\rm FT}) = {\rm Tr}'_{\cH} (-1)^F e^{-\beta \{ \cQ, \overline{\cQ} \} + 2\pi\ii \sigma_{\rm FT} (J_1 + \frac{1}{2} R) + 2\pi\ii \tau_{\rm FT} (J_2 + \frac{1}{2} R) } \,,
\end{equation}
where the trace ${\rm Tr}'$ is such that the contribution of the vacuum is 1.
Note we introduced the subscript ``FT" -- which stands for ``field theory" -- to distinguish $(\sigma_{\rm FT}, \tau_{\rm FT})$ from the ``gravitational" chemical potentials $(\sigma_g, \tau_g)$, i.e. the $(\sigma, \tau)$ we defined in \eqref{susychempot_S3}. The partition function on the background \eqref{S1S3_metric} is given by
\begin{align}\label{Zsym}
	Z_{S^1 \times S^3}(\sigma, \tau) & = {\rm Tr}_{\cH} e^{-\beta \{ \cQ, \overline{\cQ} \} + 2\pi\ii \sigma_g J_1 + 2\pi\ii \tau_g  J_2 + 2\pi\ii \Delta R} \nn \\
	& = {\rm Tr}_{\cH} e^{-\beta \{ \cQ, \overline{\cQ} \} + 2\pi\ii \sigma_g (J_1+\frac{1}{2}R) + 2\pi\ii \tau_g (J_2 + \frac{1}{2}R) + \pi \ii R } \,,
\end{align}
where the second line follows from the constraint \eqref{SYMconstraint_S3_1}. This can be put into the following form
\begin{align}\label{Zsymv2}
	Z_{S^1 \times S^3}(\sigma_g, \tau_g) & = {\rm Tr}_{\cH} e^{2\pi\ii J_1 + \pi \ii F} e^{-\beta \{ \cQ, \overline{\cQ} \} + 2\pi\ii \sigma_g (J_1+\frac{1}{2}R) + 2\pi\ii \tau_g (J_2 + \frac{1}{2}R) + \pi \ii R }\nn \\
	& = {\rm Tr}_{\cH} (-1)^F e^{-\beta \{ \cQ, \overline{\cQ} \} + 2\pi\ii (\sigma_g+1) (J_1+\frac{1}{2}R) + 2\pi\ii \tau_g (J_2 + \frac{1}{2}R)} \nn \\
	& = {\rm Tr}_{\cH} (-1)^R e^{-\beta \{ \cQ, \overline{\cQ} \} + 2\pi\ii \sigma_g (J_1+\frac{1}{2}R) + 2\pi\ii \tau_g (J_2 + \frac{1}{2}R) } \,,
\end{align}
where in the last line it is apparent that $Z_{S^1 \times S^3}(\sigma_g, \tau_g)$ is an ``index" graded by the R-charge.
This is precisely the ``index" we computed using Honda's formula in section \ref{subsection:S1S3}, that gives \eqref{roundSCI} in the Cardy-like limit.
Notice that $Z_{S^1 \times S^3}(\sigma_g, \tau_g) = \cI(\sigma_g+1,\tau_g)$, up to the normalisation of the vacuum.
Instead of starting from \eqref{Zsym} on the complex background \eqref{S1S3_metric}, with anti-periodic Killing spinors around $S^1_\beta$, one can also start from \eqref{SCIwrittenout} on a real background with periodic Killing spinors, and shift $\sigma_{\rm FT} \rightarrow \sigma_{\rm FT} + 1$ to land on the same expression, i.e. the ``second sheet" of the index \cite{Cassani:2021fyv}.

Given the constraints \eqref{SYMconstraint_S3b}, \eqref{SYMconstraint_S3v}, all of the above goes through for the $S^1_\beta \times S^3_{\mf{b}_1,\mf{b}_2}$ and $S^1_\beta \times S^3_v$ backgrounds. The chemical potentials $(\sigma,\tau)$ are simply replaced by the elliptically/biaxially squashed chemical potentials, \eqref{susychempot_S3b} and \eqref{susychempot_S3v}, respectively. The ``squashings" -- along with the ``twistings" -- enter via geometric fugacities that refine the index \cite{Aharony:2013dha,Closset:2013vra}.

Let us denote the pair of spinors associated with $\cQ$, $\overline{\cQ}$ as $\zeta_+$, $\zeta_-$.
The curved space supersymmetry algebra is given by \cite{Dumitrescu:2012ha,Klare:2012gn}
\begin{equation}
	[ \delta_{\zeta_+}, \delta_{\zeta_-} ] \Upsilon = 2 \ii \left( \cL_K - \ii r K \hook A^{\rm nm} \right) \Upsilon \,,
\end{equation}
where $\Upsilon$ is a generic field in the theory, and $r$ is its R-charge.
Given the explicit conformal Killing spinors \eqref{S1S3KSchoice}, \eqref{S1S3bKS}, and \eqref{S1S3vKS}, this can be evaluated explicitly.
For $S^1_\beta \times S^3$, it is straightforward to see that
\begin{equation}
	K \hook A^{\rm nm} = \ii \left( \Psi - \frac{3}{2} \right) \,,
\end{equation}
from \eqref{S1S3_AnmVnm} and \eqref{S1S3susyKV}. Using \eqref{AnmVnm_S1S3v}, \eqref{S1S3vsusyKV} for $S^1_\beta \times S^3_v$, and similarly \eqref{AnmVnm_S1S3b} and \eqref{S1S3bsusyKV} for $S^1_\beta \times S^3_{\mf{b}_1, \mf{b}_2}$, one finds that $K \hook A^{\rm nm}$ is unchaged for both cases.
The upshot is that the squashing only affects the coefficients in front of the angular momentum generators in the superalgebra \eqref{QQbar}.
Note the abstract operators $H$, $J_1$, $J_2$ that appear in \eqref{QQbar} are associated with $\cL_{\partial_{t_L}}$, $\cL_{\partial_{\phi_L}}$, $\cL_{\partial_{\psi_L}}$, where the $(t_L, \phi_L, \psi_L)$ coordinates are related to $(\tau, \varphi_1, \varphi_2)$ by the following coordinate transformation
\begin{equation}
	t_L = -\ii t_E \,,\qquad \phi_L = \varphi_1 - \ii \Omega_1 t_E \,,\qquad \psi_L = \varphi_2 - \ii \Omega_2 t_E \,.
\end{equation}
We thus infer that the supercharges preserved by the $S^1_\beta \times S^3_v$ background obey
\begin{equation}\label{QQbarv}
	\{ \cQ_v , \overline{\cQ}_v \} = H - \frac{1}{v} J_1 - \frac{1}{v} J_2 - \frac{3}{2} R \,.
\end{equation}
whereas those preserved by the $S^1_\beta \times S^3_{\mf{b}_1, \mf{b}_2}$ background obey
\begin{equation}\label{QQbarb}
	\{ \cQ_{\mf{b}} , \overline{\cQ}_{\mf{b}} \} = H - \mf{b}_1 J_1 - \mf{b}_2 J_2 - \frac{3}{2} R \,,
\end{equation}

It is natural to conjecture that \eqref{QQbarv}, \eqref{QQbarb} are satisfied as BPS charge relations of supersymmetric AlAdS$_5$ black hole solutions with biaxially/elliptically squashed conformal boundaries.

\section{Killing spinors on squashed three-spheres}\label{app:KS}
We collect here some useful facts about Killing spinors on round, elliptically squashed, and biaxially squashed three-spheres.

\subsection{Round \texorpdfstring{$S^3$}{S3}}
Consider $S^3$ as a surface in $\C^2$. Introducing complex coordinates $(z_1, z_2) \in \C^2$, we parametrise the unit radius $S^3$ by
\begin{equation}
	(z_1, z_2) = (\sin \vartheta e^{\ii \varphi_1}, \cos \vartheta e^{\ii \varphi_2}) \,,
\end{equation}
with $\vartheta \in [0, \pi/2]$ and $\Delta \varphi_i = 2\pi$.
The metric on $S^3$ is then given by
\begin{align}
	{\rd}s^3(S^3) & = {\rd} z_1 {\rd} \bar{z}_1 + {\rd} z_2 {\rd} \bar{z}_2 \nn \\
	& = {\rd}\vartheta^2 + \sin^2 \vartheta {\rd}\varphi_1^2 + \cos^2 \vartheta {\rd}\varphi_2^2 \,.
\end{align}

Half of the 4 Killing spinors are constant in the left-invariant frame, and are solutions to the Killing spinor equation
\begin{equation}
	\left( \nabla_a^{(3)} - \frac{\ii}{2} \gamma_a^{(3)} \right) \upsilon = 0\,,
\end{equation}
where $\gamma^{(3)}$ are elements of Cliff(3), with $a = 1,2,3$. The two spinors satisfy, respectively,
\begin{equation}\label{app:LieKS}
	\cL_{\partial_{\varphi_1}} \upsilon = \cL_{\partial_{\varphi_2}} \upsilon = c_J \frac{\ii}{2} \upsilon \,, \qquad c_J = \pm 1 \,.
\end{equation}

\subsection{Elliptically squashed \texorpdfstring{$S^3_{\mf{b}_1, \mf{b}_2}$}{S3b1b2}}
We consider elliptical squashing of the round $S^3$, specified by parameters $\mf{b}_1, \mf{b}_2$. Let
\begin{equation}
	(z_1, z_2) = \left( \frac{\sin \vartheta}{\mf{b}_1} e^{\ii \varphi_1}, \frac{\cos \vartheta}{\mf{b}_2} e^{\ii \varphi_2} \right) \,,
\end{equation}
such that
\begin{equation}
	|z_1|^2 + |z_2|^2 = \frac{\sin^2\vartheta}{\mf{b}_1} + \frac{\cos^2\vartheta}{\mf{b}_2} = 1 \,.
\end{equation}
We then obtain the metric on an ellipsoid, where
\begin{align}\label{app:ellipsoidmetric}
	{\rd}s^3(S^3_{\mf{b}_1, \mf{b}_2}) & = {\rd} z_1 {\rd} \bar{z}_1 + {\rd} z_2 {\rd} \bar{z}_2 \nn \\
	& = f(\vartheta)^2 {\rd}\vartheta^2 + \frac{\sin^2 \vartheta}{\mf{b}_1^2} {\rd}\varphi_1^2 + \frac{\cos^2 \vartheta}{\mf{b}_2^2} {\rd}\varphi_2^2 \,,
\end{align}
with
\begin{equation}
	f(\vartheta) = \sqrt{\frac{\cos^2\vartheta}{\mf{b}_1^2} + \frac{\sin^2\vartheta}{\mf{b}_2^2} } \,.
\end{equation}
As explained in \cite{Martelli:2011fu}, the second line of \eqref{app:ellipsoidmetric} is a non-singular metric on $S^3_{\mf{b}_1, \mf{b}_2}$ for any smooth function $f(\vartheta)$ of definite sign, where for regularity one should demand that $|f(\vartheta)| \rightarrow 1/\mf{b}_1$ as $\vartheta \rightarrow 0$ and $|f(\vartheta)| \rightarrow 1/\mf{b}_2$ as $\vartheta \rightarrow \frac{\pi}{2}$. The following statements hold without having to specify $f(\vartheta)$.

In alignment with conventions chosen in the main text, we shall be interested in Killing spinors on $S^3_{\mf{b}_1, \mf{b}_2}$ that are solutions to
\begin{equation}
	\left( \nabla_a^{(3)} - \ii A_a^{(3)} - \frac{\ii}{2 f(\vartheta)} \gamma_a^{(3)} \right) \upsilon = 0 \,,
\end{equation}
with the following background gauge field
\begin{equation}\label{app:S3bAfield}
	A^{(3)} = c_J \left[ \frac{1}{2} \left( 1 - \frac{1}{f(\vartheta) \mf{b}_1} \right) {\rd} \varphi_1 + \frac{1}{2} \left( 1 - \frac{1}{f(\vartheta) \mf{b}_2} \right) {\rd} \varphi_2 \right] \,.
\end{equation}
Taking the frame
\begin{equation}
	\e^1 = \frac{\sin\vartheta}{\mf{b}_1} {\rd} \varphi_1 \,,\qquad \e^2 = \frac{\cos\vartheta}{\mf{b}_2} {\rd} \varphi_2 \,,\qquad \e^3 = f(\vartheta) {\rd}\vartheta \,,
\end{equation}
and Pauli matrices for $\gamma^{(3)}_a$, the explicit solutions are given by \cite{Hama:2011ea}
\begin{equation}
	\upsilon = \frac{1}{\sqrt{2}} \begin{pmatrix}
		\ii \exp \left[ \frac{\ii}{2} (c_J (\varphi_1 + \varphi_2) + \vartheta) \right] \\
		- c_J \exp \left[ \frac{\ii}{2} (c_J (\varphi_1 + \varphi_2) - \vartheta) \right]
	\end{pmatrix} \,, \qquad c_J = \pm 1 \,.
\end{equation}
The two spinors satisfy, respectively,
\begin{equation}\label{app:LieKSb}
	\cL_{\partial_{\varphi_1}} \upsilon = \cL_{\partial_{\varphi_2}} \upsilon = c_J \frac{\ii}{2} \upsilon \,.
\end{equation}

\subsection{Biaxially squashed \texorpdfstring{$S^3_v$}{S3v}}
We consider the $SU(2) \times U(1)$--invariant biaxial (Berger) squashing of the round $S^3$, with squashing parameter $v$. The metric can be written as
\begin{align}\label{biaxsquashedS3v}
	{\rd}s^2(S^3_v) & = \frac{1}{4} \left( \sigma_1^2 + \sigma_2^2 + v^2 \sigma_3^2 \right) \nn \\
	&  = {\rd}\vartheta^2 + \sin^2\vartheta {\rd}\varphi_1^2 + \cos^2\vartheta {\rd}\varphi_2^2 + (v^2-1) \left( \sin^2\vartheta {\rd}\varphi_1 + \cos^2\vartheta {\rd}\varphi_2 \right)^2 \,,
\end{align}
where $\sigma_a$ are the standard left-invariant one-forms. The special case $v=1$ corresponds to the round $S^3$.

We shall be interested in Killing spinors that are solutions to
\begin{equation}\label{app:S3vKSE}
	\left( \nabla_a^{(3)} - \ii A_a^{(3)} - \frac{\ii}{2} v \, \gamma_a^{(3)} \right) \upsilon = 0 \,,
\end{equation}
where the background gauge field is given by
\begin{equation}\label{app:S3vAfield}
	A^{(3)} = - (v^2 - 1) \left( \sin^2\vartheta {\rd} \varphi_1 + \cos^2\vartheta {\rd} \varphi_2 \right) \,.
\end{equation}
In the left--invariant frame
\begin{align}
	\e^1 & = \frac{1}{2} \Big( -2 \sin(\varphi_1 + \varphi_2) {\rd}\vartheta + \cos(\varphi_1 + \varphi_2) \sin(2\vartheta) ({\rd}\varphi_2 - {\rd}\varphi_1) \Big) \,, \nn \\
	\e^2 & = \frac{1}{2} \Big( 2 \cos(\varphi_1 + \varphi_2) {\rd}\vartheta + \sin(\varphi_1 + \varphi_2) \sin(2\vartheta) ({\rd}\varphi_2 - {\rd}\varphi_1) \Big) \,, \nn \\
	\e^3 & = \frac{v}{2} \Big( 2 \sin^2\vartheta {\rd}\varphi_1 + 2 \cos^2\vartheta {\rd} \varphi_2 \Big) \,,
\end{align}
the two Killing spinors that solve \eqref{app:S3vKSE} are constant spinors, and they satisfy, respectively,
\begin{equation}\label{app:LieKSv}
	\cL_{\partial_{\varphi_1}} \upsilon = \cL_{\partial_{\varphi_2}} \upsilon = c_J \frac{\ii}{2} \upsilon \,, \qquad c_J = \pm 1 \,.
\end{equation}

\bibliographystyle{utphys} 
\bibliography{biblio}{}

\end{document}